\documentclass{jfm}
  \usepackage{amsmath}
  \usepackage{mathabx}
\usepackage[scr=boondoxo]{mathalfa}
  \usepackage{amssymb}
  \usepackage{gauss}
  \usepackage{bm}
  \usepackage{mwe}
  \usepackage{float}
  \usepackage{subfig}
 \usepackage{caption}
 \usepackage{float}
 \usepackage{natbib}
\usepackage{physics}
\usepackage{epstopdf, epsfig}
\usepackage{xcolor}
\usepackage{hyperref}
\usepackage{mathdots}
\usepackage{microtype}

\usepackage{leftidx}

\makeatletter
\def\sgn{\mathop{\operator@font sgn}}
\makeatother

\newcommand{\Ma}{\mbox{\it Ma}}

 \shorttitle{Effect of cavity-induced perturbation interactions}
\shortauthor{M. R. Islam, and Y. Sun}

\title{Effect of cavity-induced perturbation interactions on the transitional flow after the trailing edge}

\author{Md Rashidul Islam\aff{1}
  \corresp{\email{mislam11@syr.edu}} \and
  Yiyang Sun\aff{1}}

\affiliation{\aff{1}Department of Mechanical and Aerospace Engineering, Syracuse University, Syracuse, NY 13244, USA}
   
\begin{document}

\maketitle

\begin{abstract}
We investigate the modal and non-modal linear amplification mechanisms in the flow over a subsonic open cavity and their subsequent interactions to identify optimal flow perturbations that propagate downstream the cavity and trigger flow transitions. Using both the stationary and time-varying base flows from a Direct Numerical Simulation of a cavity flow at Mach 0.6, we employ classical and harmonic resolvent analyses to explain the role of the cavity-generated perturbations in destabilizing the flow downstream. Our analysis of perturbation amplification about the mean flow identifies structures that resemble Tollmien-Schlichting (T-S) waves at the Rossiter frequency in the attached boundary layer region after the cavity. A low-frequency centrifugal instability dominates inside the cavity. The mean flow also amplifies stationary streaks via a lift-up mechanism that extends throughout the boundary layer region downstream of the cavity. The harmonic resolvent analysis (HRA) reveals the amplification of additional perturbations by the unsteady Rossiter base flow. By restricting the input and output at the same frequency in the HRA, we find the amplification of the stationary perturbation to be the most dominant 3D instability mechanism. This amplification is driven by the interaction of the 3D streaks with the unsteady Rossiter base flow, which generates internal forcing in the form of oblique T-S waves, thereby further amplifying the streaks. The interaction between the centrifugal perturbation and the unsteady flow also generates streamwise elongated structures in the boundary layer region after the cavity. Together, the centrifugal-Rossiter and streak-Rossiter interactions synergistically amplify perturbations downstream of the cavity.
\end{abstract}



\section{Introduction}
\label{sec:introduction} 
The flow over a compressible open cavity has been a subject of numerous studies over the past decades due to its relevance in a wide range of engineering applications \citep{rowley2006dynamics}. The flow inside and surrounding the cavity exhibits rich fluid-dynamics phenomena, including flow separation, shear-layer instability, and hydrodynamic-acoustic interactions, which drive shear-layer oscillation and vortex dynamics. As a result, it also serves as a model problem for investigating these phenomena in detail, which are ubiquitous in many other flow configurations. Perhaps the most researched is how the unstable shear-layer oscillation materializes and causes intense acoustic pressure fluctuations, a phenomenon linked to the Rossiter feedback mechanism \cite{rossiter1964wind}. Recent investigations have been more focused on identifying the instability mechanism within the cavity and devising ways to suppress it \citep{liu2021unsteady,godavarthi2025phase}. A recent investigation showed that the cavity can itself serve as a passive tool for controlling flow separation on an airfoil \citep{visbal2020passive}. They found that downstream-traveling perturbations generated by the cavity can modify the separation zone and delay dynamic stall. The findings open up a question: what are the characteristics of the cavity-generated perturbations that can travel further downstream and cause flow modifications? A cavity can also represent a gap or discontinuity on the airfoil surface due to an imperfect manufacturing process, and answering this question is essential for understanding the perturbations that can cause boundary-layer transition after the gap.

Resolvent analysis has emerged as a valuable tool for understanding perturbation amplification in fluid flow due to external or internal forcing \citep{mckeon2010critical,jovanovic2005componentwise}. In the classical approach, resolvent analysis involves linearizing the Navier--Stokes equation around a stationary base flow and treating the nonlinear terms as an unknown forcing which can then be formulated in an input-output framework. The resolvent operator acts as the transfer function between the input and output perturbations in the frequency domain. The singular value decomposition (SVD) of the operator provides a way to identify the optimal perturbations that can be generated by a corresponding optimal forcing and an amplification gain between them \citep{rolandi2024invitation}. Resolvent analysis can provide insight into both the modal instabilities typically investigated in the literature using linear stability tools and the non-modal instabilities that can materialize even in the absence of any unstable eigenvalue. The analysis has been helpful for various purposes, including identifying flow instability mechanisms, designing flow control strategies, and flow estimation \citep{ribeiro2023triglobal,yeh2019resolvent,liu2021unsteady,Jung_Bhagwat_Towne_2025}

Recent extensions of the resolvent analysis modify some fundamental assumptions, thereby improving the modeling framework \citep{jovanovic2021bypass}. One such assumption is the statistical stationarity of the base flow, which limits the model's ability to predict instabilities arising from the base flow's unsteadiness. Most fluid flows are inherently unsteady, and linearizing the underlying dynamics about the stationary, e.g., the mean flow, has led to poor predictions of modal structures compared to their data-driven counterparts. Since we are interested in the perturbations carried by the unsteady cavity flow in the present study, we use the harmonic resolvent analysis \citep{padovan2020analysis,padovan2022analysis,islam2024identification}, which models the base flow as time-varying with a fundamental frequency. The analysis follows the same approach as the classical one, except for incorporating the unsteady base flow, which, in the frequency domain, results in interactions among perturbations at different frequencies through the base flow. The resulting formulation involves a modified operator, compared to the classical one, that incorporates the effects of both the mean flow and the unsteady part of the base flow. The harmonic resolvent operator acts as a transfer function between the input and the output with frequency components in a specific set. The SVD of the operator yields optimal responses with the same frequency components as the optimal forcing, and the optimal gain provides a measure of amplification \citep{padovan2022analysis,islam2024identification}.

Numerous studies have investigated the effects of cavity geometry, the upstream boundary-layer characteristics, and compressibility on instabilities within the cavity \citep{bres2008three,rowley2002self,sun2017biglobal,de2014three} using the linear stability analysis. In addition to the two-dimensional (2D) unstable Rossiter modes, the flow within the cavity is found to be susceptible to low-frequency three-dimensional (3D) instabilities. These 3D instabilities result from the centrifugal effect in the recirculating region of the flow. Recent efforts have used resolvent analysis to identify optimal locations for actuator placement to control cavity oscillations \citep{liu2021unsteady}. In analyzing the effect of the gap on boundary-layer transition, \cite{victorino2023gap} found that only in the presence of both the Rossiter and centrifugal mechanisms, with a significant growth rate identified by global stability analysis, can flow transition occur downstream. However, the role of any non-modal instability mechanism was not captured due to the nature of the stability analysis.

In this study, we treat the cavity as an amplifier of perturbations that travels downstream and triggers a flow transition after the trailing edge. The goal then is to identify the underlying mechanism that generates these perturbations and their downstream evolution by the base flow. Doing so captures the representative configurations of both the gap-induced transitional flow and the micro-cavity-generated perturbations that can passively modify downstream flow. We hypothesize that the multi-modal interactions among the cavity-generated perturbations, such as the Rossiter and centrifugal instabilities, will be crucial for explaining the downstream evolution of the flow after the cavity. Moreover, non-modal instabilities can significantly accelerate the transition process. Thus, harmonic resolvent analysis will be a powerful tool for uncovering interactions among perturbations at different frequencies and for identifying optimal modal and non-modal instabilities in the post-cavity region.

The structure of the paper is as follows. In \S\ref{sec:setup}, we discuss the governing equation and the numerical configuration used to obtain the cavity flow. We provide the theory and its numerical implementation of the tools, including linear stability, Floquet stability, and classical and harmonic resolvent analysis, which will be used comprehensively to analyze perturbation amplification in \S\ref{sec:theory}. In \S\ref{sec:sec4} we present the results of the classical resolvent analysis, which will help us identify the dominant instability mechanisms in the vicinity of the cavity. Then, in \S\ref{sec:5}, we investigate the interactions among the cavity-induced perturbations using the harmonic resolvent analysis and discuss their downstream evolution. Finally, we provide concluding remarks and future considerations in \S\ref{sec:conclusion}. In addition, we provide supplemental discussions in the appendices to complement the main text.

\section{Numerical configuration of cavity flow}\label{sec:setup}
We perform a direct numerical simulation (DNS) of the flow over a rectangular cavity with a length-to-depth ratio of $L/D=2.0$. The flow is governed by the compressible Navier-Stokes equation that reads in a compact notation
\begin{equation}
    \frac{\partial \boldsymbol{q}(\boldsymbol{x},t)}{\partial t} = \underbrace{-\nabla \cdot \boldsymbol{F}(\boldsymbol{q}(\boldsymbol{x},t))}_{\mathcal{N}(\boldsymbol{q}(\boldsymbol{x},t))},
    \label{eq2p1}
\end{equation}
where $\boldsymbol{q} = [\rho, \rho\boldsymbol{u},\rho E]^T$ is the vector of conservative variables denoting the density, momentum, and total energy of the flow, respectively, and $\boldsymbol{F}(\boldsymbol{q})$ denotes the combined Euler and viscous fluxes. The variables are made non-dimensional using the depth of the cavity $D$ as the reference length, the free-stream speed of sound $\tilde{a}_{\infty}$ as the reference velocity, the free-stream density $\tilde{\rho}_{\infty}$, and temperature $\tilde{T}_{\infty}$. We introduce the definition of the dimensionless Reynolds numbers $\Rey$, $\Rey_D$, Prandtl number $\Pran$, and Mach number $\Ma$ as
\refstepcounter{equation}
$$
  \Rey = \frac{\tilde \rho_{\infty} \tilde a_{\infty} D}{\tilde \mu_{\infty}},\quad
  \Rey_D = \frac{\tilde \rho_{\infty} \tilde U_{\infty} D}{\tilde \mu_{\infty}},\quad
  \Pran = \frac{\tilde \mu_{\infty} \tilde c_p}{\tilde \kappa},\quad
  \Ma = \frac{\tilde u}{\tilde a_{\infty}}, 
  \eqno{(\theequation{\mathit{a},\mathit{b},\mathit{c},\mathit{d}})}
$$
where $\tilde \mu_{\infty}$ is the reference dynamic viscosity, $\tilde U_{\infty}$ is the freestream velocity, $\tilde c_p$ is the specific heat capacity at constant pressure, and $\tilde \kappa$ is the thermal conductivity of the fluid. Then, the non-dimensional Navier-Stokes equation~(\ref{eq2p1}) reads in an expanded form
\begin{subequations}
\begin{equation}
\label{eq2p3a}
\frac{\partial \rho}{\partial t} + \nabla\cdot (\rho\boldsymbol{u})= 0,
\end{equation}
\begin{equation}
\label{eq2p3b}
 \frac{\partial}{\partial t} (\rho \boldsymbol u) + \nabla\cdot [\rho \boldsymbol u \otimes \boldsymbol{u} + p \boldsymbol{I} - \frac{1}{\Rey} \boldsymbol{\tau}] = 0,
\end{equation}
\begin{equation}
\label{eq2p3c}
 \frac{\partial}{\partial t} (\rho E) + \nabla\cdot [(\rho E +p)\boldsymbol u - \frac{1}{\Rey} \boldsymbol u\cdot \boldsymbol{\tau} - \frac{1}{\Rey \Pran (\gamma_s-1)}\nabla T] = 0,
\end{equation}
\end{subequations}
where $\otimes$ denote the tensor product, $\boldsymbol{\tau} = \mu[\nabla \boldsymbol{u} + (\nabla \boldsymbol{u}^T) -\frac{2}{3}(\nabla\cdot \boldsymbol u)\boldsymbol{I}]$ is the viscous stress tensor of the Newtonian fluid, $\boldsymbol{I}$ is the identity tensor, $\boldsymbol{u} = (u,v,w)^T$ denotes the velocity vector, $p$ is the pressure, $T$ is the temperature, and $E = p/[\rho(\gamma_s-1)]+\frac{1}{2} \boldsymbol{u}\cdot \boldsymbol{u}$ is the internal energy of the flow for a thermally and calorically perfect gas. To close the system of equations~(\ref{eq2p3a})-(\ref{eq2p3c}) we use the non-dimensional ideal gas law
\begin{equation}
    p = \frac{1}{\gamma_s}\rho T.
    \label{eq2p4}
\end{equation}
The dynamic viscosity is linked to the temperature according to the power law $\mu(T) = (T/T_{\infty})^{0.76}$ \citep{garnier2009large}. We assume a constant Prandtl number $\Pran=0.7$ and a specific heat ratio $\gamma_s=1.4$ for the air. 

To discretize the governing equations~(\ref{eq2p3a})-(\ref{eq2p3c}), we have used the high-fidelity computational fluid dynamics (CFD) solver CharLES \citep{bres2017unstructured}, which incorporates the second-order finite-volume method for spatial discretization. The spatial discretization of the Euler flux uses the Riemann solver at the cell faces based on the Harten-Lax-Van Leer contact (HLLC) scheme \citep{toro2019hllc}. The simulations are advanced in time using a third-order Runge-Kutta method through explicit time stepping.

\begin{figure}
  \centerline{\includegraphics{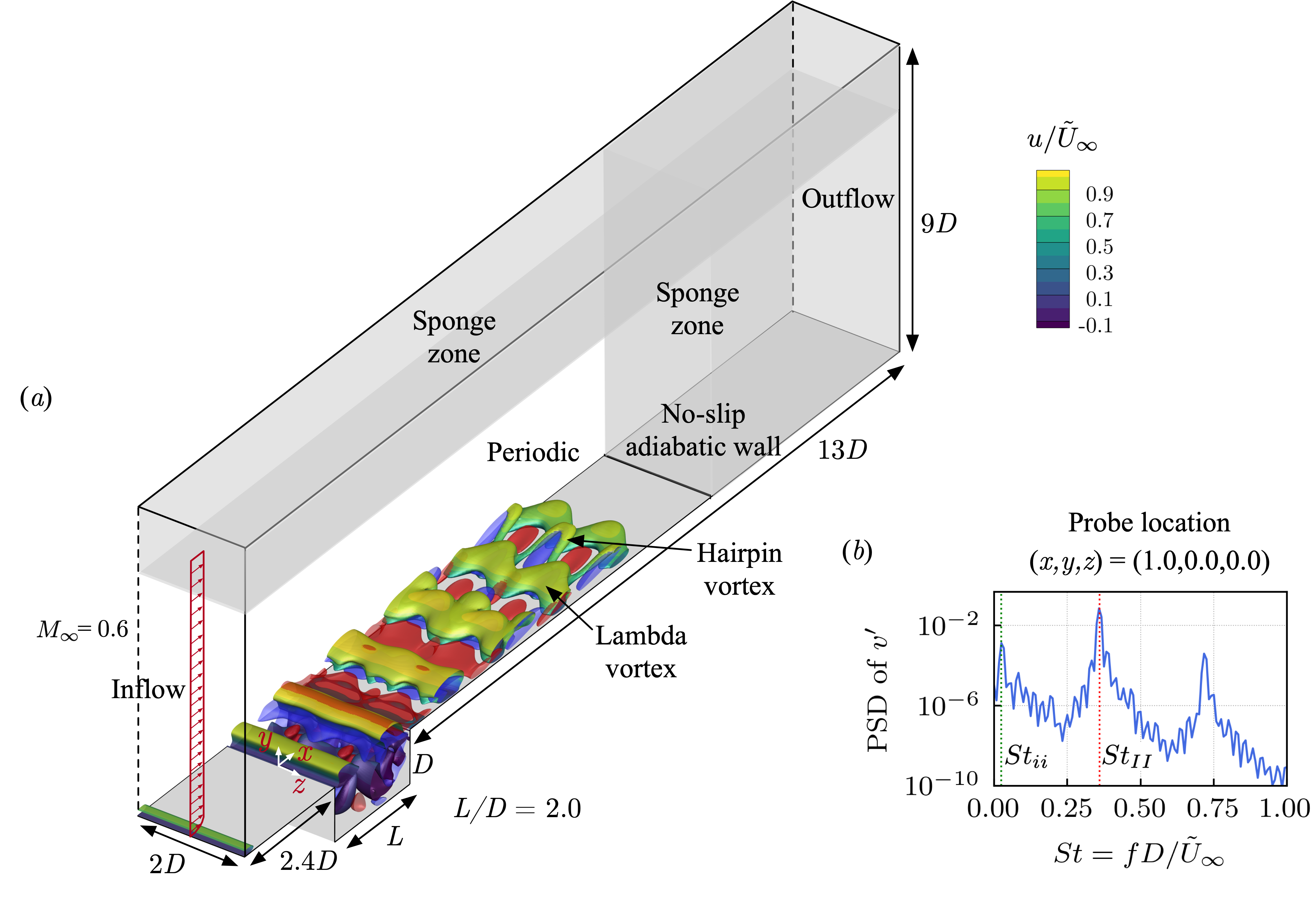}}
   \caption{(a) Computational setup for the DNS of the flow over a rectangular cavity at $\Ma_{\infty}=0.6$ (drawing is not to scale). Instantaneous iso-surface of Q-criterion at level $QD^2/\tilde U_{\infty}^2=0.03$ is colored by streamwise velocity, and iso-surface of fluctuating streamwise velocity is shown at levels $u'/\tilde U_{\infty}=\pm0.05$ (Red: positive, Blue: negative). (b) Power spectral density of the fluctuating vertical velocity ($v'$) sampled at ($x,y,z$)=(1.0,0.0,0.0).}
\label{fig:1}
\end{figure}

We have performed three-dimensional (3-D) DNS of the cavity flow at a Reynolds number based on cavity depth of $Re_D = 1500$ and free-stream Mach number of $\Ma=0.6$. The initial momentum thickness of the boundary layer at the leading edge is set to $D/\theta_0=26.4$. The velocity field in the entire domain is initialized using the incompressible Blasius velocity profile with a spreading rate in the streamwise direction, except within the cavity. The flow inside the cavity is initialized with a small Gaussian perturbation of the streamwise velocity component along the spanwise direction and zero for the other two velocity components. The schematic of the computational domain is shown in the figure~\ref{fig:1}(a). The origin of the coordinate system is located at the leading edge of the cavity and halfway along the span. The domain extends $2.4D$ upstream from the leading edge. The outflow of the domain is located at $15D$ in the streamwise direction. The top boundary of the domain is placed at a distance of $9D$ from the floor of the cavity, and the spanwise extent of the domain is $W/D=2$.

The computational domain is discretized with approximately $6.77\times 10^6$ grid points, with mesh refinement near the solid boundaries and within the cavity. At the inflow, a Blasius velocity profile is prescribed. A sponge zone \citep{freund1997proposed} is applied within a spatial extent of $5D$ from the outflow and $2D$ from the top boundary. An adiabatic no-slip condition is imposed at the cavity surface and the upstream and downstream walls. Periodic boundary conditions are applied to the spanwise boundaries.

\subsection{Flow characteristics}
A representative visualization of the instantaneous flowfield is shown in figure~\ref{fig:1}(a) using the iso-surfaces of Q-criterion at the value of $QD^2/\tilde{U}_{\infty}^2=0.03$ and colored by the streamwise velocity. The roll-up of the shear layer above the cavity and the formation of the spanwise-uniform vortical structures are evident in figure~\ref{fig:1}(a). These structures, upon impinging at the trailing edge, generate acoustic waves that travel upstream and perturb the shear layer near the leading edge, resulting in a feedback resonant loop. The power spectral density (PSD) of the vertical velocity fluctuation in figure~\ref{fig:1}(b) from a probe located in the shear layer region reveals that the dominant frequency of the resonance occurs at the Strouhal number of $St_{II} = fD/\tilde{U}_{\infty}=0.36$. The corresponding Strouhal number of $fL/\tilde{U}_{\infty}=0.72$ based on the cavity length agrees well with the Rossiter mode II frequency predicted from the semi-empirical formula \citep{rossiter1964wind}
\begin{equation}
    St = \frac{n-\xi}{M_{\infty}+1/\varphi},
    \label{Rossiter}
\end{equation}
where $n$ is the index of the Rossiter mode, the phase lag $\xi$ is 0.25, and the value of $\varphi$ is $0.57$ \citep{heller1971flow}. We also find the presence of a low-frequency peak near the Strouhal number of $St_{ii}=0.025$, which is associated with a traveling wave inside the cavity \citep{bres2008three,sun2017biglobal}. The spanwise-uniform vortical structure begins to deform near the trailing edge. After the trailing edge, the flow field resembles transitional boundary-layer flow, with the appearance of lambda vortices and the subsequent development of hairpin vortices. The question that arises is, what are the characteristics of the perturbation that convects downstream of the cavity to destabilize the flow? To answer this, we will use the classical and harmonic resolvent analyses, the theoretical formulations of which are discussed next.

\section{Description of the resolvent analysis framework}\label{sec:theory}
 In this work, we use the classical resolvent \citep{jovanovic2005componentwise,mckeon2010critical} and harmonic resolvent \citep{padovan2020analysis,padovan2022analysis,islam2024identification} analysis to model the dynamics of spanwise homogeneous small-amplitude perturbations inside and after the cavity flow. To derive the resolvent modeling framework, we decompose the conservative state variables $\boldsymbol{q}(\boldsymbol{x},t)$ as $\boldsymbol{q}(x,y,z,t) = \boldsymbol{Q}(x,y,t)+\boldsymbol{q}'(x,y,z,t)$, where $\boldsymbol{Q}(x,y,t)$ is a time-varying base flow and $\boldsymbol{q}'(\boldsymbol{x},t)$ is an infinitesimal perturbation evolving about the base flow. Substituting the decomposition into the Navier-Stokes equation~(\ref{eq2p1}) and collecting terms of the same order, we get
 \begin{equation}
     \label{eq3p1}
     \frac{\partial \boldsymbol{q}'(\boldsymbol{x},t)}{\partial t} = \underbrace{\mathcal{D}_{\boldsymbol{q}}(\mathcal{N}(\boldsymbol{Q}(x,y,t)))}_{\mathsfbi{A}(t)} \boldsymbol{q}'(\boldsymbol{x},t) + \boldsymbol{h}'(\boldsymbol{x},t),
 \end{equation}
where $\mathsfbi{A}(t)=\mathcal{D}_{\boldsymbol{q}}(\mathcal{N}(\boldsymbol{Q}(x,y,t)))$ is the linear Jacobian operator of the Navier-Stokes equation evaluated at the base flow $\boldsymbol{Q}(x,y,t)$, and $\boldsymbol{h}'(\boldsymbol{x},t)$ contains higher order terms in $\boldsymbol{q}'(\boldsymbol{x},t)$. 

\subsection{Classical resolvent analysis}
In the classical resolvent analysis, we replace the base flow $\boldsymbol{Q}(x,y,t)$ by a steady solution of equation~(\ref{eq2p1}) or a statistically stationary mean of the unsteady solution computed as 
\begin{equation}
    \label{eq3p2}
    \overline{\boldsymbol{Q}}(x,y) = \lim_{T\rightarrow\infty}\, \frac{1}{T} \int_0^T \boldsymbol{Q}(x,y,t) \mathrm{d}t,
\end{equation}
and construct the linear operator $\mathsfbi{A}(\overline{\boldsymbol{Q}})$ using the stationary state. We then express the forcing $\boldsymbol{h}'(\boldsymbol{x},t)$ and steady-state response $\boldsymbol{q}'(\boldsymbol{x},t)$ using a sum of Fourier modes as
\begin{subequations}
    \begin{equation}
    \label{eq3p3a}
     \boldsymbol{h}'(\boldsymbol{x},t)=\sum_{\beta,\omega} \hat{\boldsymbol{h}}'_{\beta,\omega}(x,y) \mathrm{e}^{\mathrm{i}(\omega t+\beta z)} + c.c.,
    \end{equation}
    \begin{equation}
    \label{eq3p3b}
     \boldsymbol{q}'(\boldsymbol{x},t)=\sum_{\beta,\omega} \hat{\boldsymbol{q}}'_{\beta,\omega}(x,y) \mathrm{e}^{\mathrm{i}(\omega t+\beta z)} + c.c.,
    \end{equation}
\end{subequations}
for real-valued frequency $\omega$ and spanwise wavenumber $\beta$, where $c.c.$ represents complex conjugate. Then, we substitute the expressions into equation~(\ref{eq3p1}) to obtain 
\begin{equation}
    \label{eqn3p4}
    \hat{\boldsymbol{q}}'_{\beta,\omega} = \mathsfbi{R}(\beta,\omega) \hat{\boldsymbol{h}}'_{\beta,\omega},
\end{equation}
where $\mathsfbi{R}(\beta,\omega)=[\mathrm{i}\omega\mathsfbi{I}-\mathsfbi{A}(\overline{\boldsymbol{Q}},\beta)]^{-1}$ is the classical resolvent operator \citep{jovanovic2005componentwise,mckeon2010critical}. Given a forcing $\hat{\boldsymbol{h}}'_{\beta,\omega}$  with the frequency $\omega$ and the spanwise wavenumber $\beta$, equation~(\ref{eqn3p4}) gives a steady state response $\hat{\boldsymbol{q}}'_{\beta,\omega}$ if the linear operator $\mathsfbi{A}(\overline{\boldsymbol{Q}},\beta)$ is stable. To check the stability of $\mathsfbi{A}(\overline{\boldsymbol{Q}},\beta)$, we consider equation~(\ref{eq3p1}) without forcing $\boldsymbol{h}'(\boldsymbol{x},t)$ and seek eigensolution of the form $\boldsymbol{q}'(\boldsymbol{x,t})=\hat{\boldsymbol{q}}(x,y)\mathrm{e}^{\lambda t +\beta z} + c.c.$, which upon substituting into equation~(\ref{eq3p1}) yields
\begin{equation}
    \label{eq3p5}
    \mathsfbi{A}(\overline{\boldsymbol{Q}},\beta) \hat{\boldsymbol{q}}(x,y) =   \lambda \hat{\boldsymbol{q}}(x,y),
\end{equation}
where $\lambda$ is the complex-valued eigenvalue. The real component of the eigenvalue $\lambda_r$ denotes the growth rate, and the imaginary component $\lambda_i$ is the temporal frequency of the eigenmode $\hat{\boldsymbol{q}}(x,y)$. If there exists an eigenvalue with $\lambda_r>0$ then $\mathsfbi{A}(\overline{\boldsymbol{Q}},\beta)$ is unstable, and the solution due to the forcing $\hat{\boldsymbol{h}}'_{\beta,\omega}$ could be obscured by the unforced dynamics of the linear system. In that case, a meaningful solution can be obtained by incorporating a discounting parameter \citep{yeh2019resolvent,jovanovic2004modeling,sun2020resolvent} $\kappa>\max(\lambda_r)$ into the resolvent operator as $\mathsfbi{R}(\beta,\omega)=[(\kappa+\mathrm{i}\omega)\mathsfbi{I}-\mathsfbi{A}(\overline{\boldsymbol{Q}},\beta)]^{-1}$. By introducing the parameter $\kappa$, we can study the linear response developing on a shorter temporal scale $1/\kappa$ than the time-scale of the dominant instabilities present in the base flow.

Following the discretization of equation~(\ref{eq3p1}), we obtain the discrete resolvent operator of dimension $\mathsfbi{R}(\beta,\omega)\in\mathbb{C}^{(5\times N)\times (5\times N)}$ with $N$ being the number of discrete grid points. In classical resolvent analysis, we are interested in finding the optimal forcing that maximizes the gain between the input and output energy. Here we measure both the input and output energy using Chu's \citep{chu1965energy,hanifi1996transient} compressible energy norm given in non-dimensional form by
\begin{equation}
    \label{eq3p6}
    ||\boldsymbol{\hat q}'_p||_E^2 = E_\text{c} = \frac{1}{2}\int_{\mathcal{V}} \boldsymbol{\hat q}_p^{' *} \text{diag}\left(\frac{\overline T}{\gamma \overline \rho},\overline \rho,\frac{\overline \rho}{\gamma(\gamma-1) \overline T}\right) \boldsymbol{\hat q}'_p \mathrm{d}\mathcal{V}\, , 
\end{equation}
where $\boldsymbol{\hat q}'_p=[\hat\rho',\hat{\boldsymbol{u}}',\hat T']$ is the vector of primitive perturbation variables. The optimization problem can be shown to result in an SVD of the weighted resolvent operator,
\begin{equation}
    \label{eq3p7}
    \mathsfbi{N}^{1/2}\mathsfbi{S}^{-1}\mathsfbi{R}(\beta,\omega)\mathsfbi{S}\mathsfbi{N}^{-1/2} = \tilde{\Psi} \Sigma \tilde{\Phi},
\end{equation}
where $\mathsfbi{N}$ is the positive definite matrix containing discrete Chu's norm weight along the main diagonal, the matrix $\mathsfbi{S}$ transforms the primitive perturbation variables to the conservative perturbation variables \citep{islam2024identification}, and $\Sigma=\text{diag}(\sigma_1,\sigma_2,\dots)$ contains the ranked singular values. The optimal energy gain is given by $\sigma_1^2$. The forcing modes at frequency $\omega$ and wavenumber $\beta$ correspond to the columns of $\Phi=\mathsfbi{N}^{-1/2}\tilde\Phi$, and the response modes at the same frequency and wavenumber are the columns of $\Psi=\mathsfbi{N}^{-1/2}\tilde\Psi$.

\subsection{Harmonic resolvent analysis}\label{sec3p2}
In harmonic resolvent analysis, we consider the base flow $\boldsymbol{Q}(x,y,t) = \boldsymbol{Q}(x,y,t+T_0)$ to be $T_0$ periodic ($T_0=2\pi/\omega_0$) and consequently, $\mathsfbi{A}(t)$ is also $T_0$-periodic. For ease of derivation, we can split the linear operator $\mathsfbi{A}(t)$ into a time-invariant operator $\mathsfbi{L}$ and a time-periodic operator $\mathsfbi{G}(t)$ that depends on $\boldsymbol{Q}(t)$ \citep{islam2024identification}. Then we expand $\mathsfbi{G}(t)$ into a Fourier series as $\mathsfbi{G}(t) = \sum_{k=-\infty}^{\infty} \mathsfbi{\skew3\hat{G}}_k \mathrm{e}^{\mathrm{i} k\omega_0 t}$. Here, we model the spanwise homogeneous forcing and the corresponding post-transient response in the frequency domain using the exponentially modulated periodic (EMP) signal \citep{wereley1990frequency} as
\begin{subequations}\label{eq3p8}
    \begin{equation}
        \boldsymbol{h}'(\boldsymbol{x},t) = \mathrm{e}^{\mathrm{i}\gamma \omega_0 t} \sum_{\beta} \sum_{m=-\infty}^{\infty} \hat{\boldsymbol{h}}'_{\beta,m}(x,y) \mathrm{e}^{\mathrm{i}(\beta z+ m\omega_0 t)},
    \end{equation}
    \begin{equation}
        \boldsymbol{q}'(\boldsymbol{x},t) = \mathrm{e}^{\mathrm{i}\gamma \omega_0 t} \sum_{\beta} \sum_{m=-\infty}^{\infty} \hat{\boldsymbol{q}}'_{\beta,m}(x,y) \mathrm{e}^{\mathrm{i}(\beta z+ m\omega_0 t)},
    \end{equation}
\end{subequations}
where $\gamma,\beta$ are real numbers and $m$ is an integer. Substituting the Fourier expansions of $\mathsfbi{G}(t), \hat{\boldsymbol{q}}'(t),$ and $\hat{\boldsymbol{h}}'(t)$ into the equation~(\ref{eq3p1}) we obtain
\begin{equation}
    \label{eq3p9}
    \mathrm{i}(\gamma+m)\omega_0 \hat{\boldsymbol{q}}'_{\beta,\gamma+m} = \mathsfbi{L}\hat{\boldsymbol{q}}'_{\beta,\gamma+m} + \sum_{l=-\infty}^{\infty} \mathsfbi{\skew3\hat{G}}_{m-l} \hat{\boldsymbol{q}}'_{\beta,\gamma+l} + \hat{\boldsymbol{h}}'_{\beta,\gamma+m},
\end{equation}
which represents a system of infinitely coupled equations where perturbation $\hat{\boldsymbol{q}}'_{\beta,\gamma+m}$ is coupled with $\hat{\boldsymbol{q}}'_{\beta,\gamma+l}$ through the base flow at frequency $(m-l)\omega_0$ for a constant spanwise wavenumber $\beta$. The infinite-dimensional coupled system of equations in (\ref{eq3p9}) can be compactly expressed in matrix form as 
\begin{equation}
    \label{eq3p10}
    [\mathrm{i}\gamma\omega_0\boldsymbol{I}-\mathsfbi{T}(\beta)] \hat{\mathcal{Q}}_{\beta,\gamma} = \hat{\mathcal{F}}_{\beta,\gamma},
\end{equation}
where

\begin{equation}
    \label{eq3p11a}
    \setlength{\arraycolsep}{4pt}
    \renewcommand{\arraystretch}{1.5}
    \mathsfbi T(\beta) = \left[
    \begin{array}{ccccc}
     \ddots & \vdots& \vdots& \vdots& \iddots  \\
      \displaystyle
      \dots& \mathsfbi R_{-1}^{-1}& \mathsfbi{\skew3\hat{G}}_{-1}& \mathsfbi{\skew3\hat{G}}_{-2} &\dots  \\
      \displaystyle
      \dots& \mathsfbi{\skew3\hat{G}}_1 & \mathsfbi R_{0}^{-1}& \mathsfbi{\skew3\hat{G}}_{-1} &\dots\\
      \displaystyle
      \dots& \mathsfbi{\skew3\hat{G}}_2& \mathsfbi{\skew3\hat{G}}_1 & \mathsfbi R_{1}^{-1} &\dots\\
      \displaystyle
      \iddots & \vdots& \vdots& \vdots& \ddots
    \end{array}  \right] ,
    \stepcounter{equation} \tag*{(\theequation{\textit{a}})}
\end{equation}
\begin{equation}
    \label{eq3p11b}
    \setlength{\arraycolsep}{4pt}
    \renewcommand{\arraystretch}{1.4}
    \hat{\mathcal Q}_{\beta,\gamma} = \left[
    \begin{array}{c}
     \vdots\\ \hat{\boldsymbol q}'_{\beta,\gamma-1}\\ \hat{\boldsymbol q}'_{\beta,\gamma}\\ \hat{\boldsymbol q}'_{\beta,\gamma+1}\\ \vdots
    \end{array}  \right], \quad
    \setlength{\arraycolsep}{4pt}
    \renewcommand{\arraystretch}{1.4}
    \hat{\mathcal F}_{\beta,\gamma} = \left[
    \begin{array}{c}
     \vdots\\ \hat{\boldsymbol h}'_{\beta,\gamma-1}\\ \hat{\boldsymbol h}'_{\beta,\gamma}\\ \hat{\boldsymbol h}'_{\beta,\gamma+1}\\ \vdots
    \end{array}  \right],
    \tag*{(\theequation{\textit{b},\textit{c}})}
\end{equation}
and $\mathsfbi{R}^{-1}_k = (-\mathrm{i}k\omega_0\boldsymbol{I}+\mathsfbi{L}+\mathsfbi{\skew3\hat{G}}_0)$. We note that the operator $\mathsfbi{T}$ can be singular if linearization is done around an unforced periodic base flow $\boldsymbol{Q}(t)$ \citep{padovan2022analysis,islam2024identification}, and projecting the dynamics onto a subspace that eliminates the singularity might be required to avoid numerical difficulties and get physically interesting solutions \citep{padovan2022analysis}. We then define the harmonic resolvent operator as $\mathsfbi{H}(\beta,\gamma)=[\mathrm{i}\gamma \omega_0\mathsfbi{I}-\mathsfbi{T}(\beta)]^{-1}$ that acts on any given input $\mathcal{F}_{\beta,\gamma}$ to give the steady state output $\mathcal{Q}_{\beta,\gamma}$ if the linear operator $\mathsfbi{T}(\beta)$ is stable. To check the stability of the system, we again consider the unforced dynamics of equation~(\ref{eq3p1}) with an eigensolution of the form $\boldsymbol{q'}(\boldsymbol{x},t)=\tilde{\boldsymbol{q}}_{\beta}(x,y,t)\mathrm{e}^{s t+\mathrm{i}\beta z} + c.c.$ to obtain
\begin{equation}
    \label{eq3p12}
     \frac{\partial \tilde{\boldsymbol{q}}_{\beta}(x,y,t)}{\partial t} + s \tilde{\boldsymbol{q}}_{\beta}(x,y,t) = \underbrace{\mathsfbi{A}(t)}_{\mathsfbi{L}+\mathsfbi{G}(t)}\tilde{\boldsymbol{q}}_{\beta}(x,y,t),
\end{equation}
where $s \in \mathbb{C}$ is the Floquet exponent. We express the $T_0$-periodic Floquet mode $\tilde{\boldsymbol{q}}_{\beta}(x,y,t)$ and the linear operator $\mathsfbi{G}(t)$ using their respective Fourier series expansions $\tilde{\boldsymbol{q}}_{\beta}(x,y,t)=\sum_{m=-\infty}^{\infty} \hat{\boldsymbol{q}}_{\beta,m}(x,y)\mathrm{e}^{\mathrm{i}m\omega_0 t}$ and $\tilde{\mathsfbi{G}}(x,y,t)=\sum_{m=-\infty}^{\infty} \hat{\mathsfbi{G}}_{m}(x,y)\mathrm{e}^{\mathrm{i}m\omega_0 t}$, and substituting into equation~(\ref{eq3p10}), we get the following infinite-dimensional matrix eigenvalue problem \citep{wereley1991linear,franceschini2022identification}
\begin{equation}
    \label{eq3p13}
    \mathsfbi{T}(\beta)\hat{\mathcal{Q}}_{\beta} = s \mathsfbi{I}
    \hat{\mathcal{Q}}_{\beta},
\end{equation}
which results from the so-called Floquet-Hill theory. The infinite-dimensional operator $\mathsfbi{T}(\beta)$ is known as the Hill matrix, which has the same expression as equation~(\ref{eq3p9}). The infinite-dimensional eigenvector associated with $s$ is 
\begin{equation}
    \label{eq3p14}
    \hat{\mathcal{Q}}_{\beta}^T = \begin{bmatrix} \dots& \hat{\boldsymbol{q}}_{\beta,-1}&\hat{\boldsymbol{q}}_{\beta,0}&\hat{\boldsymbol{q}}_{\beta,1}& \dots\end{bmatrix}.
\end{equation}
If the real component of any Floquet exponent $\mathrm{Re}(s)>0$, then the linear system is unstable. In that case, we introduce a discounting parameter $\kappa$ into the harmonic resolvent operator as $\mathsfbi{H}(\beta,\gamma)=[(\kappa+\mathrm{i}\gamma \omega_0)\mathsfbi{I}-\mathsfbi{T}(\beta)]^{-1}$ to study the forced dynamics over a shorter timescale similar to strategy used in the the classical resolvent analysis with a discounting parameter.

\subsection{Selection of frequency set for harmonic resolvent analysis}
Although theoretically, we can assume an infinite number of harmonics for the Fourier expansions of the base flow $\boldsymbol{Q}(t)$ and perturbations $\boldsymbol{q}'(t), \boldsymbol{h}'(t)$, in practical computation, we truncate them to a finite size to keep the computational expense tractable. Henceforth, we will assume a finite set of base flow frequencies $k\omega_0\in\Omega_b=\{-N_b,\dots,-1,0,1,\dots,N_b\}\omega_0$ with $2N_b+1$ harmonics for the Fourier expansion of $\boldsymbol{Q}(t)$ and consequently of $\boldsymbol{G}(t)$. For perturbations, we will consider a finite set of frequencies $\Omega_{\gamma} = \gamma \omega_0+ m\omega_0$ with $m\omega_0\in\{-N_h,\dots,-1,0,1,\dots,N_h\}\omega_0$ resulting in $2N_h+1$ harmonics for the expansions of $\boldsymbol{q}'(t)$ and $\boldsymbol{h}'(t)$ in equation~(\ref{eq3p8}). For fluid flow, $\boldsymbol{Q}(t)$ can usually be approximated by some dominant large-scale motion in the flow. It can be represented in the frequency domain with its angular frequency $\omega_0$ and a few harmonics yielding $N_b\leq N_h$. The final problem involves an operator $\mathsfbi{T}(\beta)$ with dimension $N_p\times N_p$ where $N_p=5\times N\times(2N_h+1))$ and $N$ is the number of discrete grid points.

In harmonic resolvent analysis, we seek the input $\hat{\mathcal{F}}_{\beta,\gamma}$ over the set of temporal frequencies $\Omega_{\gamma}$ and constant wavenumber $\beta$ that maximize the energetic response $\hat{\mathcal{Q}}_{\beta,\gamma}$ over the same set of frequencies and the wavenumber. The optimization problem can be shown to result in an SVD of the weighted harmonic resolvent operator \citep{islam2024identification}, giving
\begin{equation}
    \label{eq3p15}
    \mathsfbi{W}^{1/2}\mathsfbi{M}^{-1}\mathsfbi{H}(\beta,\gamma)\mathsfbi{M}\mathsfbi{W}^{-1/2} = \tilde{\mathsfbi{U}}\Sigma \tilde{\mathsfbi{V}}^*, 
\end{equation}
where the matrix $\mathsfbi{W}= \mathrm{blkdiag}\{\mathsfbi{N}\}$ contains the Chu's norm weights along the main diagonal, matrix $\mathsfbi{M}$ converts the primitive perturbation variables to conservative form in the frequency domain \citep{islam2024identification}, and $\Sigma=\text{diag}(\sigma_1,\sigma_2,\dots)$ contains the ranked singular values. The forcing modes at the set of temporal frequencies $\Omega_{\gamma}$ and wavenumber $\beta$ corresponds to the columns of $\mathsfbi{V}=\mathsfbi{W}^{-1/2}\tilde{\mathsfbi{V}}$ and response mode at the same frequencies and wavenumber are the columns of $\mathsfbi{U}=\mathsfbi{W}^{-1/2}\tilde{\mathsfbi{U}}$. 

\subsection{Numerical computation of the resolvent and stability problems}{\label{sec3p4}}
To compute the base flow $\boldsymbol{Q}(t)$ about which linearizations are done, we advanced the DNS described in \S\ref{sec:setup} in time until the initial transient period is over. We collected $N_s=3000$ snapshots of spanwise-averaged flow state variables $\boldsymbol{q}(x,y,t)$ spanning over a non-dimensional period of $t\tilde U_{\infty}/D=120$. We take the dominant frequency of oscillation in the cavity, i.e., the angular frequency $\omega_0 = (2\pi \tilde{U}_{\infty}/D)St_{II}$ of the Rossiter mode II, as the fundamental frequency for approximating the base flow $\boldsymbol{Q}(t)$. Then we evaluated the Fourier coefficients of the conservative state variables $\hat{\boldsymbol{Q}}_k$ at frequencies $k\omega_0$ as 
\begin{equation}
    \label{eq3p16}
    \hat{\boldsymbol{Q}}_k(x,y) = \frac{1}{N_s} \sum_{j=0}^{N_s-1} \boldsymbol{q}(x,y,j) \mathrm{e}^{\mathrm{i}k\omega_0 j\Delta t},
\end{equation}
where $k\omega_0\in\Omega_b=\{-2,-1,0,1,2\}\omega_0$ are the set of base flow frequencies. A set of perturbation frequencies $\Omega_{\gamma}$ with harmonics up to $N_h=5$ will be considered. For $k=0$ in equation~(\ref{eq3p16}), we obtained the span-averaged, time-averaged base flow $\overline{\boldsymbol{Q}}$, which is used in the classical resolvent analysis.

We have interpolated the base flow data onto a computational grid with streamwise extent $x/D \in [-2.4, 14]$ and wall-normal extent $y/D \in [-1, 8]$ to generate the linear operators. Approximately $N= 53,000$ grid points are used to discretize the domain with mesh refinement near the wall and cavity. We imposed a no-slip adiabatic condition for the perturbations, i.e., $\hat{\boldsymbol{u}}'\cdot \boldsymbol{n}=0$ where $\boldsymbol{n}$ is the unit normal vector at the surface and $\nabla T'=0$, at the walls and cavity surface. A sponge zone within a $1D$ extent from the inflow, $2D$ from the top boundary, and $3D$ from the outflow is applied to bring the perturbation values to zero.

The discretized linear operator for the classical resolvent analysis has a size of the order $\mathcal{O}(10^5)$. For the harmonic resolvent analysis, the size of the linear operator is of order $\mathcal{O}(10^6)$. Thus, to efficiently compute the SVD of such large operators, we use a randomized algorithm \citep{ribeiro2020randomized}. The algorithm was implemented in an in-house solver that uses the PETSc \citep{balay2019petsc} and SLEPc \citep{hernandez2005slepc} libraries for parallelization. The linear equations in the algorithm are solved directly via a parallel LU decomposition of the sparse linear operators distributed across multiple CPUs using the MUMPS library \citep{amestoy2000mumps}. We have used $10$ random test vectors for the SVD calculation. To compute the eigenspectrum of $\mathsfbi{A}(\overline{\boldsymbol{Q}},\beta)$ and $\mathsfbi{T}(\beta)$, we used a shift-invert strategy along with the Krylov-Schur method based on the PETSc and SLEPc libraries, with the dimension of the Krylov subspace being set to 24.

\section{Optimal perturbations about the mean flow}\label{sec:sec4}
\begin{figure}
  \centerline{\includegraphics{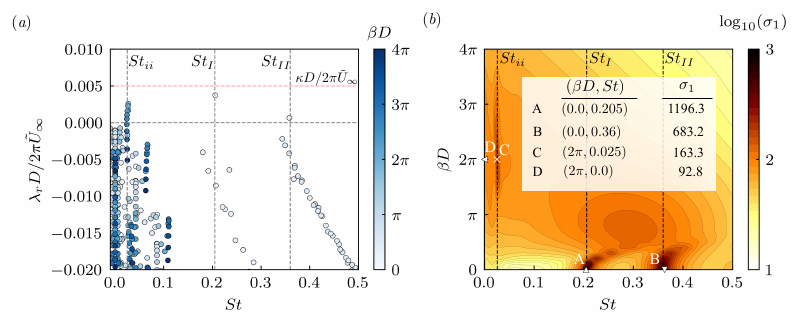}}
   \caption{(a) Eigenspectrum of the operator $\mathsfbi{A}(\overline{\boldsymbol{Q}},\beta)$. (b) The optimal energy gain $\sigma_1$ over ranges of spanwise wavenumbers ($\beta D$) and frequencies ($St$) about the spanwise and time-averaged base flow.}
\label{fig:2}
\end{figure}

We begin our analysis by performing a classical resolvent analysis, considering the spanwise and time-averaged flow $\overline{\boldsymbol{Q}}(x,y)$ as the base state. The results are crucial for understanding the origin and nature of the different linear instability mechanisms surrounding the cavity. Meanwhile, the results will serve as a benchmark for the forthcoming analysis, which will involve disturbance evolution around a time-varying base flow.
As a precursor to the resolvent analysis, we first need to perform the global stability of the linear operator $\mathsfbi{A}(\overline{Q},\beta)$ to determine whether any discounting configuration is required due to the presence of unstable modes \citep{yeh2019resolvent,liu2021unsteady}. The real and imaginary components of the non-dimensional eigenvalues of the linear operator for different spanwise wavenumbers are shown in figure~{\ref{fig:2}}(a).

The real component $\lambda_r D/2\pi \tilde U_{\infty}$ defines the growth/decay rate for values greater/less than zero, and the imaginary component $St=\lambda_i D/2\pi \tilde U_{\infty}$ denotes the frequency of the eigenmode. For 2D perturbations ($\beta D=0$), we find the presence of an unstable eigenmode with frequency $St_I=0.206$, which is associated with the Rossiter mode I \citep{sun2017biglobal} with growth rate of $\lambda_r D/2\pi\tilde{U}_{\infty}=0.0037$. We also identify a 2D unstable mode at the frequency of the Rossiter mode II ($St_{II}=0.36$) in figure~\ref{fig:2}(a) with a growth rate of $\lambda_r D/2\pi\tilde{U}_{\infty}=0.0006$, which is the dominant frequency of oscillation present in the nonlinear flow simulation. At frequencies less than approximately an order of magnitude of the Rossiter modes, we can locate an unstable branch of eigenvalues around $St_{ii}=0.025$. These eigenmodes are unstable for the spanwise wavenumber within the range $\beta D\in[1.5,2.75]\pi$ and are associated with the centrifugal instability mechanism in the literature \citep{bres2008three,sun2017biglobal}. Since the operator $\mathsfbi{A}(\overline{Q},\beta)$ is unstable, we choose to perform resolvent analysis with the discounting parameter $\kappa D/2\pi \tilde U_{\infty}=0.005$, which satisfies the criterion that the discounting parameter is greater than the maximum non-dimensional growth rate of all the eigenvalues. 

We then conduct a parametric study to identify the frequency and spanwise wavenumber of the dominant instabilities linearly amplified by external or intrinsic forcing using the resolvent analysis, considering a pair of frequencies and spanwise wavenumbers within the ranges $St\in[0,0.5]$ and $\beta D\in[0,4\pi]$, respectively. In figure~\ref{fig:2}(b), we show the contour map of the square root of the optimal gain, that is, the first singular value ($\sigma_1$) of the resolvent operator, calculated for different pairs of wavenumbers and frequencies $(\beta D,St)$. We denote several noticeable disturbance amplification regions in the map and provide the corresponding value of $\sigma_1$ in figure~\ref{fig:2}(b). Based on their location in ($\beta D,St$) space, we can identify three different classes of disturbance amplification:
\begin{enumerate}
    \item 2D ($\beta D=0$) Rossiter modes at frequencies $St_I = 0.205$ (denoted by $A$) and $St_{II}=0.36$ (denoted by $B$). Among them, we can identify the presence of frequency $St_{II}$ in the nonlinear flow, but the disturbance amplification at $St_{I}$ is not apparent when no specific external forcing is present in the flow.
    \item Amplification of 3D ($\beta D\neq 0$) traveling wave instabilities at $(\beta D,St)=(2.0\pi,0.025)$ (denoted by $C$), which we can distinctly locate in the frequency spectrum of the nonlinear flow in figure~\ref{fig:1}(b). 
    \item 3D ($\beta D\neq 0$) stationary ($St=0$) perturbation amplification (denoted by $D$). The optimal gains of this instability class are prominent in the same wavenumber range as the traveling wave with frequency $St_{ii}=0.025$. However, for a particular wavenumber in $\beta D\in[1.5,2.74]\pi$, these steady perturbations are suboptimal to the traveling wave perturbations.
\end{enumerate}

\begin{figure}
  \centerline{\includegraphics[width=0.99\textwidth]{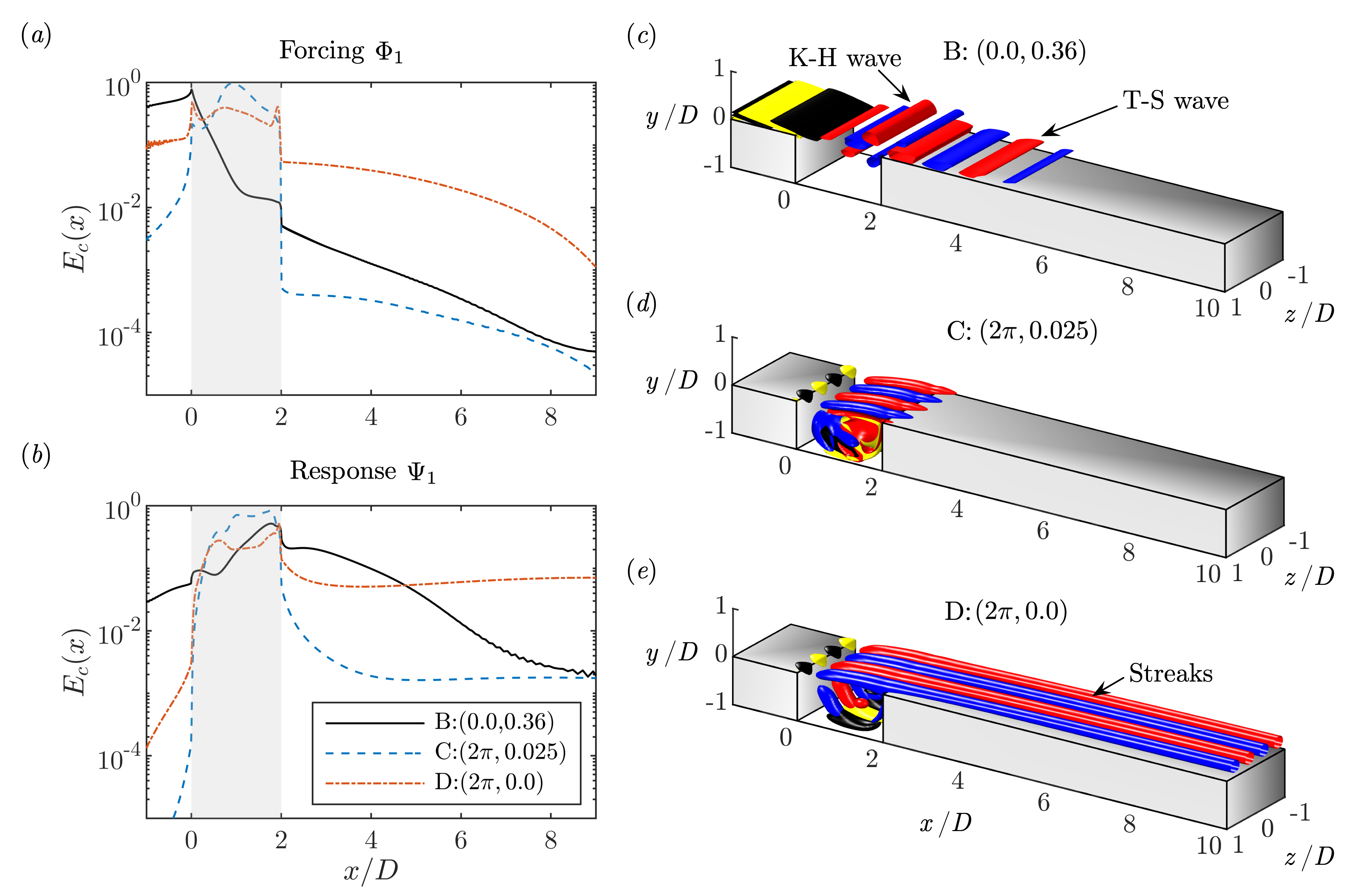}}
   \caption{(a,b) Streamwise variation of the disturbance energy of the optimal forcing and response modes and (c,d,e) iso-surface of the real component of the optimal forcing and response modes obtained from CRA at ($\beta D, St$)=$(0.0,0.36),(2\pi,0.025),(2\pi,0.0)$. The gray solid region in (a,b) indicates the extent of the cavity. The forcing modes are plotted at levels $\pm0.2\max(|\mathcal{R}\{\phi_{u}\}|)$ (Yellow: positive, Black: negative), and the response modes are plotted at levels $\pm0.3\max(|\mathcal{R}\{\psi_{u}\}|)$ (Red: positive, Blue: negative) in (c,d,e).} 
\label{fig:3}
\end{figure}

\begin{figure}
  \centerline{\includegraphics{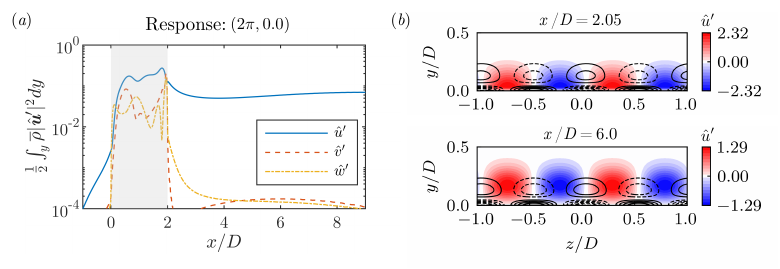}}
   \caption{(a) Streamwise variation of the disturbance energy in three velocity components of the optimal classical resolvent response mode at $(\beta D,St)=(2\pi,0.0)$. (b) Planar ($y$--$z$) view of the spatial structures of the optimal streamwise velocity response (contour plots) and streamwise vorticity forcing (contour lines, solid-positive, dashed-negative) corresponding to $(\beta D,St)=(2\pi,0.0)$.} 
\label{fig:4}
\end{figure}

To understand the underlying physical mechanism of the identified instability waves, we look into the streamwise variation of the integrated Chu's energy density of the optimal forcing and response modes in figure~\ref{fig:3}(a) obtained using 
\begin{equation}
    E_c(x) = \frac{1}{2} \int_y \left[\frac{\overline T}{\gamma \overline \rho} |\hat{\rho}'|^2 + \overline \rho(|\hat{u}'|^2+|\hat{v}'|^2+|\hat{w}'|^2)+\frac{\overline \rho}{\gamma(\gamma-1) \overline T}|\hat{T}'|^2\right] \mathrm{d}y.
\end{equation}
The integration of $E_c(x)$ along the streamwise direction results in unity by construction. The spatial mode shapes of the optimal forcing/response are also shown alongside in figure~\ref{fig:3}(c-e).
The forcing energy of the planar instability wave, denoted by $B$ with $(St,\beta D)=(0.0,0.36)$, is mostly concentrated upstream of the cavity leading edge. The energy in response gradually increases across the cavity length, indicating a convective nature of perturbation amplification. The response modal structures at $(St,\beta D)=(0.0,0.36)$ in figure~\ref{fig:3}(c) show that the instability mechanism in the shear-layer region is associated with the Kelvin-Helmholtz (K-H) waves that propagate downstream. These K-H waves interact with the upstream propagating acoustic waves inside the cavity to form an unstable Rossiter mode. Downstream of the trailing edge, the modal structures near the surface resemble the Tollmien–Schlichting (T-S) wave typically found in the wall-bounded shear flows \citep{schmid2012stability}. The amplification of T-S waves is facilitated by the Orr mechanism \citep{orr1907stability,symon2018non}, which is evident from the tilting of the forcing and response structures against and along the mean shear direction, respectively. The secondary instability of the T-S waves is one of the different ways that can lead to the laminar-to-turbulent flow transition in the boundary layer \citep{kachanov1984resonant,herbert1988secondary}. The forcing and response energy of the traveling wave, denoted by $C$ with $(St,\beta D)=(2.0\pi,0.025)$, are confined within the extent of the cavity, indicating that the instability is of absolute type. This is also evident from the forcing and response mode structures in figure~\ref{fig:3}(d). The response mode structures are located in the recirculating flow region inside the cavity, and the instability mechanism is due to a centrifugal effect \citep{bres2008three}. Another class of instability, which is stationary, has most of its energy in forcing located within the cavity as shown for $(St,\beta D)=(2.0\pi,0.0)$ in figure~\ref{fig:3}(a). The response energy of this perturbation is amplified within the cavity and persists after the trailing edge. As evident in figure~\ref{fig:3}(e), the response modal structures at $(St,\beta D)=(2.0\pi,0.0)$ are present both in the shear layer and inside the cavity. The elongated structures after the trailing edge resemble streaks observed in wall-bounded shear flows such as the Blasius boundary layer \citep{monokrousos2010global,rigas2021nonlinear}. The existence of a stationary ($St=0.0$) branch of eigenvalues has been noted in the works of \cite{bres2008three}, \cite{de2014three}, the activity of which is mainly confined within the cavity. To verify whether the structures after the trailing edge are streaks, we further look into the streamwise variation of the disturbance energy in three velocity components of the optimal response in figure~\ref{fig:4}(a). We observe that the response energy after the trailing edge is dominated by the streamwise velocity component ($\hat{u}'$) compared to the other two components ($\hat{v}',\hat{w}'$), which is a salient characteristic of streaks \citep{monokrousos2010global,schmid2002stability}. The generation of streaks through a linear non-modal lift-up mechanism \citep{landahl1980note,symon2018non} involves optimal forcing in the form of streamwise rolls (i. e. vorticity). To figure out whether such a mechanism is at play for the generation of the streaks, we show, in figure~\ref{fig:4}(b), the contours of the streamwise velocity response superimposed with the streamwise vorticity forcing in the $y$--$z$ plane at the streamwise locations $x/D=2.05$ and 6.0. At both locations, we observe that the streamwise-elongated streaky perturbations are induced by counter-rotating streamwise vortices located between alternating high-speed and low-speed streamwise velocity perturbations, confirming the amplification mechanism as the lift-up effect.

In summary, the classical resolvent analysis reveals that the planar waves ($\beta D=0$) at the Rossiter mode frequencies are the most amplified perturbations about the mean flow. The perturbations take the form of K-H instability in the separated shear layer over the cavity and T-S wave in the boundary layer after the trailing edge. A modal 3D traveling wave perturbation is present within the cavity due to a centrifugal instability. Near the cavity trailing edge, a non-modal lift-up mechanism generates stationary, streaky perturbations that exist after the trailing edge. In \S\ref{sec:5} we will study the amplification of perturbations evolving around a time-dependent base flow.

\section{Optimal perturbations about the time-varying base flow}\label{sec:5}
In this section, we apply the harmonic resolvent analysis to understand the effects of base-flow unsteadiness on the amplified perturbations within and after the cavity via different linear mechanisms. To perform the harmonic resolvent analysis, we first compute the 2-D time-varying base flow from the spanwise-averaged temporal flow snapshots of the DNS using the method discussed in \S\ref{sec3p4}. Note that we will use the nondimensional frequency expressed using the Strouhal number $St=\omega D/2\pi \tilde{U}_{\infty}$ instead of the angular frequency $\omega$ throughout this section. Thus the set of frequencies $\Omega_b= \{-2,-1,0,1,2\}St_{II}$, where $St_{II}=0.36$ is the Rossiter mode frequency, constitute the base flow $\boldsymbol{Q}(t)$. The set of perturbation frequencies considered here is $\Omega_{St_{\gamma}} = St_{\gamma} + \{-5,\dots,-1,0,1,\dots,5\}St_{II}$, where $St_{\gamma}=\gamma \omega_0 D/2\pi \tilde{U}_{\infty}$ is the center frequency in the set.

\begin{figure}
  \centerline{\includegraphics{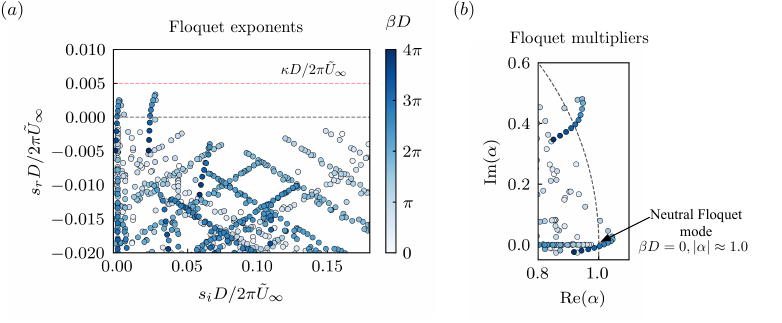}}
   \caption{(a) Normalized Floquet exponents $s$ of the operator $\mathsfbi{T}(\beta)$ in the complex plane for $s_iD/2\pi \tilde{U}_{\infty}\in [0,St_{II}/2]$. (b) A zoomed plot of Floquet multipliers $\alpha=\mathrm{e}^{sT_0}$ showing the growth of perturbations over the period $T_0$.}
\label{fig:5}
\end{figure} 
\subsection{Floquet stability analysis}
We perform a stability analysis of the operator $\mathsfbi{T}(\beta)$ that precedes the harmonic resolvent analysis to determine whether a discounting parameter is needed, as in the classical resolvent analysis, due to the presence of unstable eigenvalues. Since the base flow is time-periodic with period $T_0=D/(St_{II} \tilde{U}_{\infty})$, we apply the Floquet analysis to analyze the stability of $\mathsfbi{T}(\beta)$. In particular, we use the Floquet-Hill theory outlined in \S\ref{sec3p2} to formulate the stability analysis in the frequency domain, thereby reducing the problem to solving the eigenvalue problem~(\ref{eq3p13}). In figure~\ref{fig:5}(a), we show the complex-valued Floquet exponents, which are the eigenvalues of the operator $\mathsfbi{T}(\beta)$ obtained for a range of non-dimensional spanwise wavenumber ($\beta D$). We remark here that the Floquet exponents repeat along the imaginary axis at an interval of $St_{II}$, that is, the Floquet exponents on a section of $s_iD/2\pi \tilde{U}_{\infty}\in [(2k-1)St_{II}/2,(2k+1)St_{II}/2)$, where $k$ is an integer, uniquely represent the spectrum of $\mathsfbi{T}(\beta)$ \citep{franceschini2022identification,wereley1991linear}. However, this property might break towards the far end along the imaginary axis due to the truncation of the Fourier expansions of the base flow and perturbations using a finite number of harmonics. Moreover, theoretically (strictly for $\beta D=0$), the Floquet exponents $s$ have the same value for $\pm s_iD/2\pi \tilde{U}_{\infty}$, i.e., they are complex conjugate pairs. Here we consider $k=0$, and show the Floquet spectrum within the interval $s_iD/2\pi \tilde{U}_{\infty}\in[0,St_{II}/2]$ in the figure~\ref{fig:5}(a). We find the presence of two unstable ($s_rD/2\pi \tilde{U}_{\infty}>0.0$) branches of Floquet exponents located around $s_iD/2\pi \tilde{U}_{\infty}=0.0$ and $0.025$. Equivalently, we can also analyze the stability using the Floquet multipliers $\alpha=\mathrm{e}^{sT_0}$, and the magnitude ($|\alpha|$) of which reveals the growth/decay of perturbations over one period ($T_0$) of the base flow. The Floquet multipliers in figure~\ref{fig:5}(b) again show the two unstable branches of the Floquet mode with $|\alpha|>1.0$, indicating these modes will grow over many periods of the base flow. We find these unstable modes to be present within the spanwise wavenumber range of approximately $\beta D\in[1.5,3.0]\pi$. For 2D perturbations ($\beta D=0$), all the modes are stable. However, since the periodic base flow in the present case develops without the influence of external forcing, the leading mode at $\beta D=0$ would correspond to the zero Floquet exponent ($s=0$), or equivalently Floquet multiplier with $|\alpha|=1.0$, and the eigenmode would align with the time derivative of the base flow $d\hat{\boldsymbol{Q}}/dt$. For this neutral Floquet mode denoted in figure~\ref{fig:5}(b), we obtained $|\alpha|\approx0.99$ and the weighted projection coefficient between the corresponding eigenmode and $d\hat{\boldsymbol{Q}}/dt$ to be 0.95. Since the operator $\mathsfbi{T}(\beta)$ is unstable, we modify the harmonic resolvent operator as $\mathsfbi{H}(\beta,\gamma)=[(\kappa+\mathrm{i}\gamma \omega_0)\mathsfbi{I}-\mathsfbi{T}(\beta)]^{-1}$, where we choose the value of discounting to be $\kappa D/2\pi \tilde{U}_{\infty}=0.005$, which is higher than the non-dimensional growth rate of all the Floquet exponents.

\subsection{Harmonic resolvent analysis}
\begin{figure}
  \centerline{\includegraphics[width=0.99\textwidth]{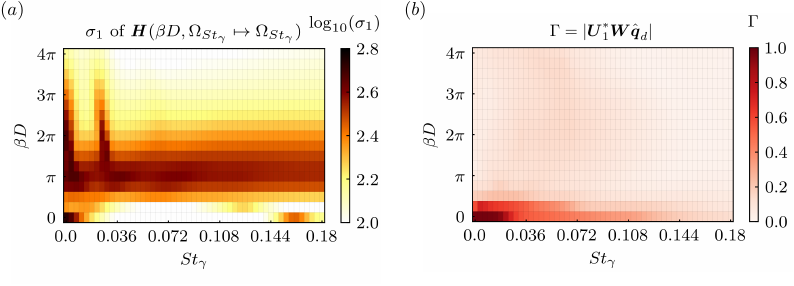}}
   \caption{(a) The optimal energy gain of the perturbation about the time-varying base flow over ranges of wavenumbers ($\beta D$) and a set of frequencies $\Omega_{St_{\gamma}}$. (b) Projection coefficient between the optimal response $U_1$ of $\mathsfbi{H}(\beta D,\Omega_{St_{\gamma}}\mapsto \Omega_{St_{\gamma}})$ and the time-derivative of base flow $d\hat{\boldsymbol{Q}}/{dt}$.}
\label{fig:6}
\end{figure}

We conduct the discounted harmonic resolvent analysis over a range of spanwise wavenumbers $\beta D\in[0,4\pi]$ with an interval of $0.25\pi$. Similar to the Floquet exponents, the singular values of the harmonic resolvent operator are also periodic at an interval of $St_{II}$, and, thus, we vary the $St_{\gamma}$ in the range $[0,St_{II}/2]$ for a particular wavenumber to obtain the unique singular values. The distribution of the first singular values ($\sigma_1$) of $\mathsfbi{H}(\beta D,\Omega_{St_{\gamma}}\mapsto \Omega_{St_{\gamma}})$ for the pairs of wavenumber ($\beta D$) and frequency set ($\Omega_{St_{\gamma}}$) are shown in figure~\ref{fig:6}(a).
Here $\Omega_{St_{\gamma}}\mapsto \Omega_{St_{\gamma}}$ indicates that $\mathsfbi{H}$ maps an input with frequency content in the set $\Omega_{St_{\gamma}}$ to an output with the same set of frequencies. Then, unlike classical resolvent analysis, here $\sigma_1$ represents the square root of the optimal gain between input and output comprised of frequencies in the set $\Omega_{St_{\gamma}}$ rather than a single frequency ($St$). For planar wave perturbations ($\beta D=0$), we obtain the highest gain for the frequency set with center frequency $St_{\gamma}=0$. This gain is due to the neutral Floquet mode that aligns with $d\hat{\boldsymbol{Q}}/dt$, which can be excluded by incorporating a projection operator during the computation of SVD \citep{padovan2022analysis}. Here we use the projection coefficient $\Gamma = |U_1^* \boldsymbol{W} \hat{q}_d|$ between the optimal response $U_1$ and a unit norm vector ($||\hat{q}_d||_{\boldsymbol{W}}=1.0$) along $d\hat{\boldsymbol{Q}}/dt$ as \textit{a posterior} measure to identify the response that aligns with the phase shift direction ($d\hat{\boldsymbol{Q}}/dt$). We show the resulting projection coefficients ($\Gamma$) in figure~\ref{fig:6}(b), where we find the optimal response at $\beta D=0$ and $St_{\gamma}\in[0,0.05]$ to have $\Gamma>0.9$ indicating a significant alignment of these modes along $d\hat{\boldsymbol{Q}}/dt$. Since the phase-shift direction is uninteresting, we exclude these response modes, along with the associated forcing and gain, from the discussion. The most amplified 3D perturbations ($\beta D\neq 0$) belong to the frequencies in the set with center frequency $St_{\gamma}\rightarrow 0$. These perturbations at $St_{\gamma}=0$ are amplified over a range of wavenumbers, with an increasing optimal gain that attains a maximum at $\beta D=2\pi$, then decreases. Because the optimal gain computed this way corresponds to a set of frequencies $\Omega_{St_{\gamma}}$, it is not comparable to the optimal gain from the classical resolvent analysis. Also, the forcing that generates different instability mechanisms in the fluid flow might not be well represented by a superposition of frequencies in a set. To remedy this, we can examine the gain obtained by varying the configuration of the input and output frequency components within the set $\Omega_{St_{\gamma}}$. The intuitive choice, which leads to a direct comparison with the classical resolvent gain, is to restrict the input and output frequency components to the same harmonic that we will analyze next. 

\subsection{Frequency restricted HRA}
\begin{figure}
  \centerline{\includegraphics{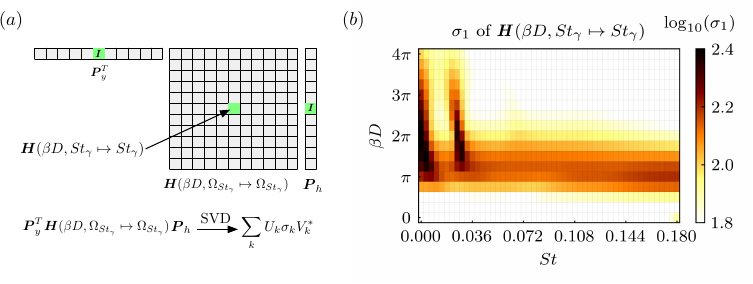}}
   \caption{(a) Schematic showing the method used for obtaining the gain between the input-output at the same frequency from the harmonic resolvent operator $\mathsfbi{H}(\beta D,\Omega_{St_{\gamma}}\mapsto \Omega_{St_{\gamma}})$ by introducing the frequency prolongation and restriction operator $\mathsfbi{P}_h$ and $\mathsfbi{P}_y$, respectively. (b) Optimal singular value distribution in the frequency-wavenumber space of the operator $\mathsfbi{H}(\beta D,St_{\gamma}\mapsto St_{\gamma})$, which maps the input-output at the same frequency component.}
\label{fig:7}
\end{figure}

To achieve the objective of getting the output at the same frequency component as the input, we introduce a frequency prolongation operator $\mathsfbi{P}_h = \text{diag}(\mathsfbi{0},\dots,\mathsfbi{I},\dots,\mathsfbi{0})\in \mathbb{R}^{N_p\times(5\times N)}$ and a frequency restriction operator $\mathsfbi{P}_y = \text{diag}(\mathsfbi{0},\dots,\mathsfbi{I},\dots,\mathsfbi{0})^T\in \mathbb{R}^{(5\times N)\times N_p}$ where $(\cdot)^T$ defines matrix transposition. The operator $\mathsfbi{P}_h$ takes an input of size $(5\times N)\times1$ and puts it as the center frequency ($St_{\gamma}$) component in an augmented vector of size $N_p \times1$ comprising of all the frequency components in a set $\Omega_{St_{\gamma}}$. Conversely, the operator $\mathsfbi{P}_y$ extracts the output at the center frequency component from a vector of size $N_p\times 1$. Then the modified operator $\mathsfbi{H}(\beta D,St_{\gamma}\mapsto St_{\gamma})$ that maps input-output at the same frequency content can be obtained from the harmonic resolvent operator after multiplying by $\mathsfbi{P}_y^T$ and $\mathsfbi{P}_h$ from left and right, respectively, as shown in figure~\ref{fig:7}(a).
The operator $\mathsfbi{H}(\beta D,St_{\gamma}\mapsto St_{\gamma})$ is also known as the mean resolvent operator \citep{leclercq2023mean,bongarzone2025mean}, which provides the mean linear response of a time-varying base flow to forcing in the statistically steady regime. Whereas in CRA, we compute the optimal linear time-invariant (LTI) input-output dynamics about a time-invariant base flow, the forcing and response of $\mathsfbi{H}(\beta D,St_{\gamma}\mapsto St_{\gamma})$ provides an optimal LTI approximation of the input-output dynamics of time-varying flows in the statistically steady regime \citep{leclercq2023mean}. We compute the SVD of the operator $\mathsfbi{H}(\beta D,St_{\gamma}\mapsto St_{\gamma})$ using the same method described in \S\ref{sec3p4}. The resulting distribution of the optimal singular values in the frequency-wavenumber $(St-\beta D)$ space is plotted in figure~\ref {fig:7}(b). Here, we show the temporal frequencies of the input in the range $St\in[0.0,0.18]$, which allows us to study the mean response of the unsteady base flow evolving at the Rossiter frequency of 0.36 to low-frequency perturbations. Within this temporal frequency range, the 3D perturbations ($\beta D\neq0$) are the most amplified. In particular, the perturbation amplifications are most pronounced at the temporal frequencies of $0.0$ and $0.025$ and in the immediate vicinity of them, for a range of spanwise wavenumbers $\beta D\in[\pi,3\pi]$. We find that the maximum gain occurs at the spanwise wavenumber $\beta D=2\pi$ for both temporal frequencies. We choose this spanwise wavenumber of $2\pi$ to investigate further the amplification characteristics around the stationary and time-varying base flows. 

\begin{figure}
  \centerline{\includegraphics[width=0.99\textwidth]{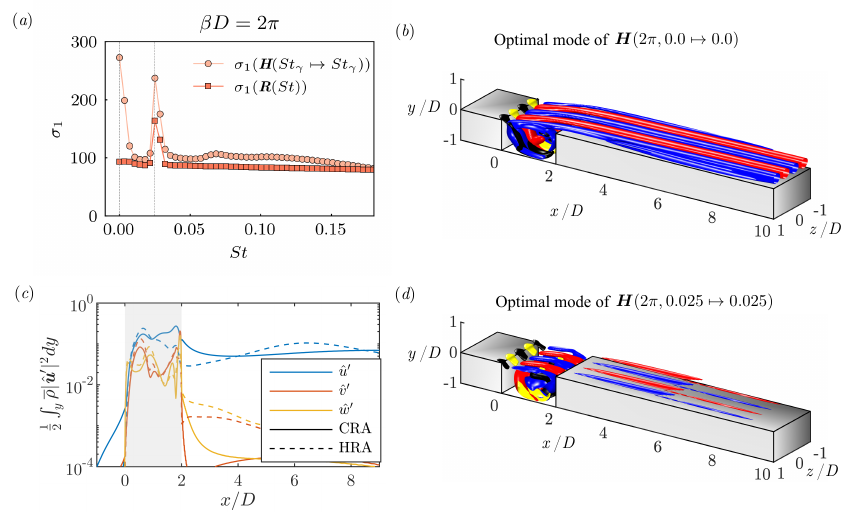}}
   \caption{(a) Comparison of the optimal gain as a function of temporal frequencies $St\in[0.0,0.18]$ at the spanwise wavenumber $2\pi$ obtained using the CRA and frequency-restricted HRA. (b,d) Optimal forcing and response mode from the frequency-restricted HRA at temporal frequencies $St=0.0$ and $0.025$. The forcing modes are plotted at levels $\pm20\%$ of the maximum absolute value (Yellow: positive, Black: negative), and the response modes are plotted at levels $\pm30\%$ of the maximum absolute value (Red: positive, Blue: negative). (c) Comparison of the streamwise variation of energy density between the optimal velocity response of the CRA and frequency-restricted HRA at $(2\pi,0.0)$.}
\label{fig:8}
\end{figure}

We compare the optimal gain obtained from the CRA and frequency-restricted HRA within the frequency range $St\in[0.0,0.18]$ at the spanwise wavenumber $2\pi$ in figure~\ref{fig:8}(a). Both analyses, using the stationary and time-varying base flows, respectively, show perturbation amplification at the centrifugal mode, as indicated by the peak at $St =0.025$. However, the higher gain of the unsteady base flow indicates stronger perturbation amplification relative to the amplification of perturbation about the stationary base flow in CRA. Furthermore, the stationary ($St=0.0$) perturbation amplification about the unsteady base flow is conspicuous from the presence of a distinct peak, which is absent in the gain obtained using the stationary base flow. The apparent discrepancy between the gains at both frequencies can be attributed to additional perturbation amplifications arising from the interaction with the unsteady part of the base flow evolving at the Rossiter frequency and its harmonics. 

The real component of the optimal streamwise velocity forcing and response modes at the frequency $St=0.0$ computed from the frequency-restricted HRA is displayed in figure~\ref{fig:8}(b). For this stationary perturbation ($St=0.0$), we observe the presence of streamwise elongated structures extending from the cavity towards the downstream region, similar to the CRA mode observed in figure~\ref{fig:3}(e). Some difference in the modal structures arises due to the consideration of the unsteady base flow, which can be quantitatively examined from the streamwise variation of the response energy of the three velocity components shown in figure~\ref{fig:8}(c). We still observe an order-of-magnitude separation between the streamwise velocity response energy and the energy in the other two components after the trailing edge from the HRA mode, suggesting the structures are streaks. However, a noticeable decrease in energy in the $\hat{u}'$ component of the frequency-restricted HRA response in the vicinity of the cavity trailing edge can be observed, followed by a gradual increase that saturates around $x/D=6.0$ in comparison to the CRA mode.

\begin{figure}
  \centerline{\includegraphics{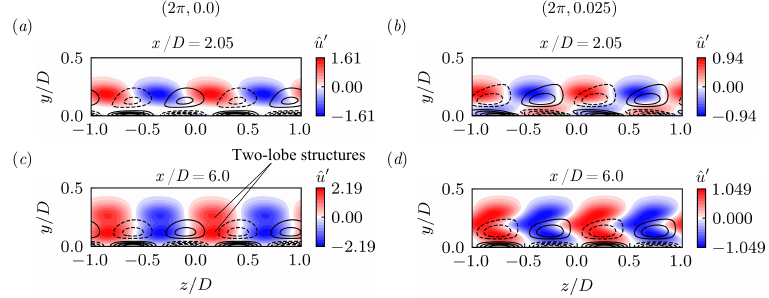}}
   \caption{Planar ($y$--$z$) view of the spatial structures of the optimal streamwise velocity response (contour plots) and streamwise vorticity forcing (contour lines, solid-positive, dashed-negative) computed from the frequency-restricted HRA corresponding to (a,c) $(\beta D,St)=(2.0\pi,0.0)$ and (b,d) $(\beta D,St)=(2.0\pi,0.025)$.}
\label{fig:9}
\end{figure}

Furthermore, we look into the $y$--$z$ planar view of the stationary ($St=0.0$) streamwise velocity response, along with the superimposed streamwise vorticity forcing at the locations $x/D =2.05$ and $6.0$ in figure~\ref{fig:9}(a,c). The forcing and response show a spatial phase shift along the spanwise direction, leading to the counter-rotating vortices being sandwiched between the velocity response. However, the structures are slightly inclined along the $y$ direction compared to the CRA modes observed in figure~\ref{fig:4}(b). In the immediate vicinity of the trailing edge ($x/D=2.05$), the prominent response structures are elevated from the wall in the cross-stream direction ($y$) in contrast to the CRA mode. Further downstream, the response structures at $x/D=6.0$ in the cross-stream direction exhibit two maxima, resulting in a two-lobe structure, a feature not observed in the CRA mode. As we will discuss shortly, the harmonic resolvent operator combines the perturbation interaction with the mean flow and subsequent interactions with the unsteady part of the base flow, the latter of which induces cross-frequency interactions that drive the observed changes in the modal structures.

The optimal mode of the frequency-restricted harmonic resolvent operator at the frequency of the centrifugal instability ($St=0.025$) is shown in figure~\ref{fig:8}(d). Within the cavity, the modal structures are located along the recirculating region driven by the centrifugal instability. After the trailing edge, streamwise elongated structures are observed near the boundary layer, which were absent in the amplified perturbations about the mean flow in CRA. The modal structures in the $y$--$z$ plane after the trailing edge at the streamwise locations $x/D=2.05$ and $x/D=6.0$ are plotted in figure~\ref{fig:9}(b,d) to show the streamwise development of the perturbations. Near the trailing edge at $x/D=2.05$, there is a spatial phase shift between the modal structures in the top and bottom layers along the $z$-direction. Unlike the stationary perturbations, the counter-rotating streamwise vorticity forcing is not located precisely between the high-speed and low-speed velocity perturbations. At $x/D=6.0$, the velocity perturbations in the upper and bottom layers are tilted against each other along the $y$-direction. Next, we will address the question of how these amplified perturbations arise after the trailing edge due to the interaction of the low-frequency centrifugal perturbation with the unsteady Rossiter base flow.

\subsection{Role of unsteady Rossiter base flow in perturbation amplification}
\begin{figure}
  \centerline{\includegraphics[width=0.95\textwidth]{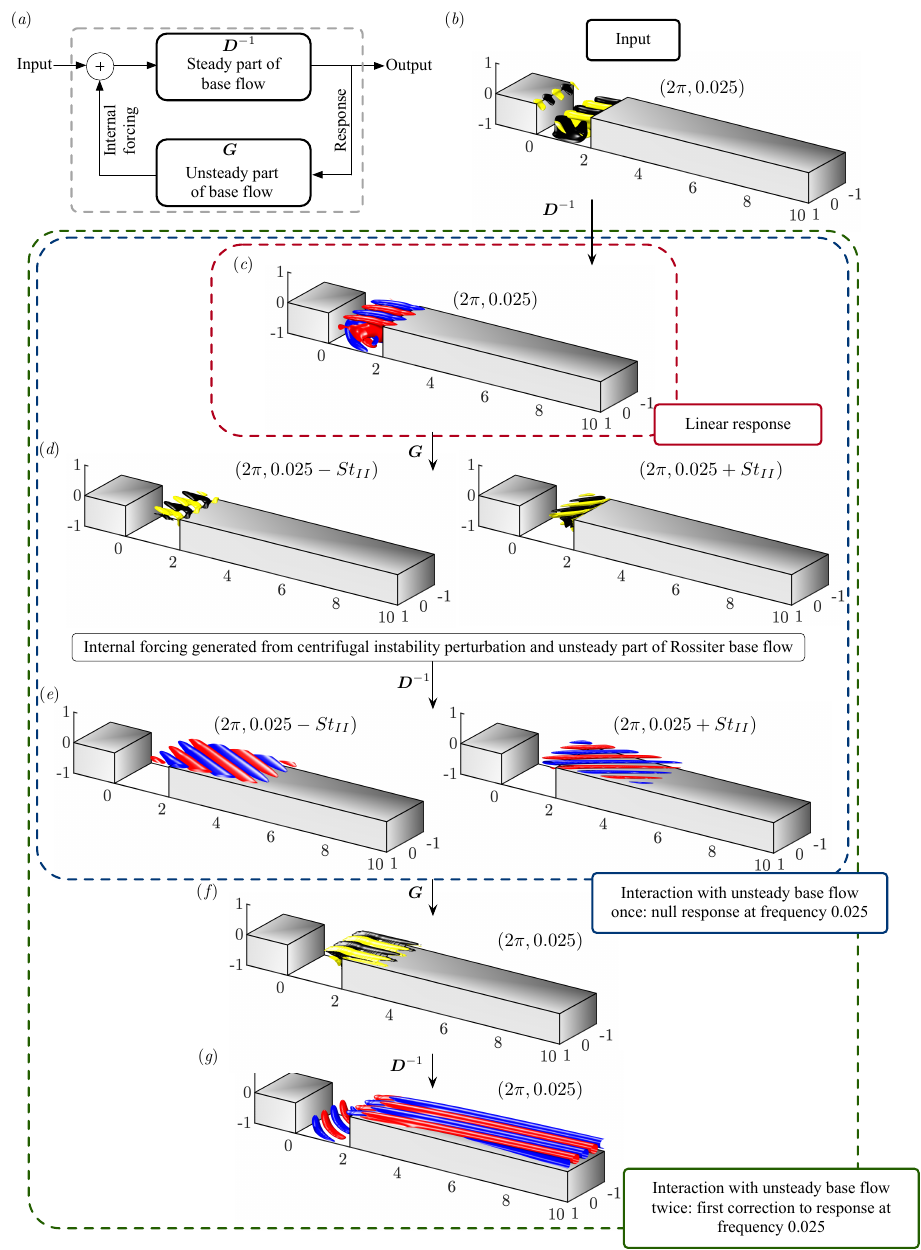}}
   \caption{(a) Block diagram representation of the operator $\mathsfbi{G}$ as a feedback loop to the main block, consisting of the operator $\mathsfbi{D}$. Sequence of perturbation generation by interaction with the steady and the unsteady part of the harmonic resolvent operator, (b,c) streamwise velocity input and corresponding response at $(2\pi,0.0)$, (d,e) internal forcing (spanwise velocity) at $(2\pi,0.025-St_{II})$ and $(2\pi,0.025+St_{II})$ generated from first interaction with unsteady part of the flow and corresponding response, (f,g) internal forcing (spanwise velocity) at $(2\pi,0.0)$ generated from second interaction with unsteady feedback loop and the associated response by the steady part of base flow. }
\label{fig:10}
\end{figure}

To understand how the perturbations get amplified due to the interaction with the unsteady part of the base flow, we will look into the structure of the inverse of the harmonic resolvent operator [$\mathrm{i}\gamma\omega_0\mathsfbi{I}-\mathsfbi{T}$], where the expression for $\mathsfbi{T}$ is given in equation~\ref{eq3p11a}. We follow the approach of \cite{leclercq2023mean} and decompose the operator as 
\begin{equation}
    \mathrm{i}\gamma\omega_0\mathsfbi{I}-\mathsfbi{T} = \mathsfbi{D}-\mathsfbi{G},
\end{equation}
where $\mathsfbi{D}$ contains the main diagonal blocks of the operator [$\mathrm{i}\gamma\omega_0\mathsfbi{I}-\mathsfbi{T}$] and the off-diagonal blocks are contained in $\mathsfbi{G}$. The expressions of the operators $\mathsfbi{D}$ and $\mathsfbi{G}$ are given in Appendix~\ref{appA}. The main diagonal elements of the operator $\mathsfbi{D}$ are the inverse of the resolvent operator constructed around the mean Jacobian of the unsteady base flow. Mathematically, for incompressible flow, the Jacobian about the mean flow $\mathsfbi{A}(\overline{\boldsymbol{Q}})$ and the mean Jacobian $\mathsfbi{L}+\hat{\mathsfbi{G}}_0$ are equivalent due to the quadratic nonlinearity. In compressible flow, the nonlinearities are cubic, and, technically, they are not equivalent. However, as we will see later, the amplification of perturbation about the Jacobian evaluated at the mean flow and the mean Jacobian are almost the same in the present study, since the unsteady part of the base flow is relatively weak compared to the mean component. These diagonal blocks in $\mathsfbi{D}$ define the amplification of perturbation due to the interaction with the steady component of the flow. The off-diagonal blocks in $\mathsfbi{G}$ are the Jacobian evaluated at the non-zero frequency component, i.e., the unsteady component of the base flow. Then the harmonic resolvent operator can be expressed as
\begin{equation}
    \underbrace{(\mathsfbi{D}-\mathsfbi{G})^{-1}}_{\mathsfbi{H}} = (\mathsfbi{I}-\mathsfbi{D}^{-1}\mathsfbi{G})^{-1}\mathsfbi{D}^{-1},
    \label{eq4p2}
\end{equation}
where the right-hand side can be presented as a feedback loop between $\mathsfbi{D}^{-1}$ and $\mathsfbi{G}$ in the block diagram representation, as shown in figure~\ref{fig:10}(a). The operator $\mathsfbi{D}^{-1}$ forms the main loop, which takes an input and provides an output amplified by the diagonal blocks that define the mean-flow resolvent operators at the frequencies in $\Omega_{St_\gamma}$. The output is then multiplied by $\mathsfbi{G}$ in the feedback loop, leading to cross-frequency interactions. Following \citep{leclercq2023mean} we can then express the operator $\mathsfbi{H}$ as a sum of the following series
\begin{equation}
\label{eq:5p3}
    \underbrace{(\mathsfbi{I}-\mathsfbi{D}^{-1}\mathsfbi{G})^{-1}\mathsfbi{D}^{-1}}_{\mathsfbi{H}} = \mathsfbi{D}^{-1} + \mathsfbi{D}^{-1} \mathsfbi{G} \mathsfbi{D}^{-1} + \mathsfbi{D}^{-1} \mathsfbi{G} \mathsfbi{D}^{-1} \mathsfbi{G} \mathsfbi{D}^{-1} + \dots,
\end{equation}
where terms on the right-hand side capture the increasing number of interactions with the unsteady part of the base flow through the feedback loop. The first term has zero interactions, the second term models the effects of interacting once with the unsteady part, and so on. For the frequency-restricted HRA, we multiply the terms on the right-hand side by the operators $\mathsfbi{P}_y^t$ and $\mathsfbi{P}_h$ from left and right, respectively. 

We then take the optimal forcing at the $(\beta D,St)=(2\pi,0.025)$ obtained from the frequency-restricted HRA and compute the output by allowing different numbers of interactions with the unsteady part of the base flow. The outputs resulting from varying numbers of interactions with unsteady flow are shown in figure~\ref{fig:10}(b-g) following the feedback mechanism described above. The sequence of generation of the outputs is as follows:
\begin{enumerate}
    \item The input first interacts with the steady part of the base flow (multiplication by $\mathsfbi{D}^{-1}$) to generate the response at $(\beta D,St)=(2\pi,0.025)$ shown in figure~\ref{fig:10}(c). We find that this output modal structure is identical to that obtained from the classical resolvent about the mean flow. Even the ratio between the norm of the output and the input results in approximately the same gain ($\sigma_1$) obtained about the mean flow. 

    \item The response at frequency 0.025 generated by the steady component then interacts with the unsteady base flow in the feedback loop (multiplication by $\mathsfbi{G}$), generating internal forcing at frequencies $0.025-St_{II}$, $0.025+St_{II}$, and so on due to cross-frequency interactions by the off-diagonal elements of $\mathsfbi{G}$. The spanwise velocity structures in figure~\ref{fig:10}(d) show that the forcing generated by the centrifugal-Rossiter interaction is concentrated near the cavity trailing edge. However, multiplication by $\mathsfbi{G}$ once does not generate any internal forcing at the center frequency component, i.e., at 0.025. The forcing then gets amplified by the steady flow component through the operator $\mathsfbi{D}^{-1}$, resulting in responses being generated at all frequency components in the set $\Omega_{St_{\gamma}}$ excluding the center frequency component. Thus, a single interaction with the unsteady feedback loop does not modify the response of the frequency-restricted HRA. The representative response structures at the frequencies $0.025-St_{II}$ and $0.025+St_{II}$ are visualized in figure~\ref{fig:10}(e).

    \item The interaction between the unsteady feedback loop and the output from the previous step now generates an internal forcing at the center frequency 0.025. The spanwise velocity component of this forcing is plotted in figure~\ref{fig:10}(f). The amplification of the forcing by the steady base flow component then generates perturbation structures after the trailing edge, as evident in figure~\ref{fig:10}(g). 
\end{enumerate}
The sequence continues for higher-order interactions with the unsteady flow. The final output of the frequency-restricted HRA is a result of the superposition of such outputs obtained by multiple interactions with the unsteady base flow. 

We can also understand the higher amplification gain of the stationary perturbation ($St=0.0$) and the associated change in the modal structures obtained in the frequency-restricted HRA through a similar process. In particular, linear response at the stationary perturbation generates internal forcing at the oblique T-S waves, i.e., at ($\beta D,St$)=($2\pi,-St_{II}$) and ($2\pi,St_{II}$) through first interaction with the unsteady Rossiter flow. Subsequent interactions with the mean and unsteady flows amplify the streaks at frequency $St=0.0$. Since the gain obtained from zero interaction with the unsteady flow matches that of the CRA, the increase in the gain is due to the amplification of the streak by the unsteady flow further downstream of the cavity. 

\section{Conclusion}\label{sec:conclusion}
The focus of this study is to uncover the origin and characteristics of the cavity-generated perturbations that are present in the transitional flow after the cavity. The perturbation amplifications are analyzed using classical and harmonic resolvent analysis, obtained by linearizing the flow dynamics about a spanwise-time-averaged flow and a spanwise-averaged unsteady flow oscillating at the Rossiter mode frequency present in the DNS. The stationary flow of the cavity is globally unstable to the 2D Rossiter mechanism and a low-frequency 3D centrifugal mechanism. The classical resolvent analysis also identified perturbation amplification due to those modal instabilities. In addition, the classical resolvent analysis predicts non-modal amplification of the stationary streaks, originating near the cavity trailing edge and extending downstream of the cavity. The most prominent modal structures that are present after the cavity are the 2D Rossiter and 3D streaky perturbations, as revealed by the optimal response of the classical resolvent analysis. In contrast, the activity of the centrifugal instability is mainly confined within the cavity and in the vicinity of the trailing edge. 

Harmonic resolvent analysis is employed for the first time to understand the role of base-flow unsteadiness in amplifying spanwise-homogeneous 3D perturbations. The results from the harmonic resolvent analysis, obtained by restricting the input-output at the same frequency, predict that the unsteady Rossiter base flow amplifies the 3D steady streaky perturbations the most. Another notable amplification occurs at the frequency of the 3D centrifugal instability, leading to the generation of streamwise elongated structures downstream of the cavity, unlike amplification about the stationary flow. By recasting the unsteady part of the linearized Jacobian as a feedback loop to the steady part, we then modeled the origin of the additional perturbation that arises after the trailing edge. We find that the unsteady base flow generates internal forcing through interactions with the perturbations amplified by the steady component of the flow. The subsequent response to the internal forcing is amplified by the steady flow, generating perturbations downstream of the cavity. The findings not only deepen understanding of traditional cavity studies but also expand the knowledge for potential uses and applications of cavity flow that leverage its downstream flow patterns.

\section*{Acknowledgments}
This material is based upon work supported by the Air Force Office of Scientific Research under award number FA9550-24-1-0136 (Program Officer: Dr.~Gregg Abate). This work used the Anvil supercomputer at Purdue University through allocations PHY250015 and PHY250306 from the ACCESS program, which is supported by U.S. National Science Foundation. We also acknowledge the computing resources provided by Syracuse University Research Computing.

\appendix
\section{Expression of the operator $\mathsfbi{D}$ and $\mathsfbi{G}$}\label{appA}
\begin{equation}
    \setlength{\arraycolsep}{4pt}
    \renewcommand{\arraystretch}{1.4}
    \mathsfbi D = \left[
    \begin{array}{ccccc}
     \ddots & \vdots& \vdots& \vdots& \iddots  \\
      \displaystyle
      \dots& \mathrm{i}\gamma\omega_0\mathsfbi{I}-\mathsfbi R_{-1}^{-1}& \mathsfbi{0}& \mathsfbi{0} &\dots  \\
      \displaystyle
      \dots& \mathsfbi{0} & \mathrm{i}\gamma\omega_0\mathsfbi{I}-\mathsfbi R_{0}^{-1}& \mathsfbi{0} &\dots\\
      \displaystyle
      \dots& \mathsfbi{0}& \mathsfbi{0} & \mathrm{i}\gamma\omega_0\mathsfbi{I}-\mathsfbi R_{1}^{-1} &\dots\\ 
      \displaystyle
      \iddots & \vdots& \vdots& \vdots& \ddots
    \end{array}  \right],
\end{equation}

\begin{equation}
    \setlength{\arraycolsep}{4pt}
    \renewcommand{\arraystretch}{1.4}
    \setlength{\arraycolsep}{4pt}
    \renewcommand{\arraystretch}{1.4}
    \mathsfbi G = \left[
    \begin{array}{ccccc}
     \ddots & \vdots& \vdots& \vdots& \iddots  \\
      \displaystyle
      \dots& \mathsfbi{0}& \mathsfbi{\skew3\hat{G}}_{-1}& \mathsfbi{\skew3\hat{G}}_{-2} &\dots  \\
      \displaystyle
      \dots& \mathsfbi{\skew3\hat{G}}_1 & \mathsfbi{0}& \mathsfbi{\skew3\hat{G}}_{-1} &\dots\\
      \displaystyle
      \dots& \mathsfbi{\skew3\hat{G}}_2& \mathsfbi{\skew3\hat{G}}_1 & \mathsfbi{0} &\dots\\
      \displaystyle
      \iddots & \vdots& \vdots& \vdots& \ddots
    \end{array}  \right].
\end{equation}

\bibliographystyle{jfm}
\bibliography{jfm}

@article{schmid2002stability,
  title={Stability and transition in shear flows. applied mathematical sciences, vol. 142},
  author={Schmid, Peter J and Henningson, Dan S and Jankowski, DF},
  journal={Appl. Mech. Rev.},
  volume={55},
  number={3},
  pages={B57--B59},
  year={2002}
}

@article{bres2008three,
  title={Three-dimensional instabilities in compressible flow over open cavities},
  author={Bres, Guillaume A and Colonius, Tim},
  journal={Journal of Fluid Mechanics},
  volume={599},
  pages={309--339},
  year={2008},
  publisher={Cambridge University Press}
}

@article{symon2018non,
  title={Non-normality and classification of amplification mechanisms in stability and resolvent analysis},
  author={Symon, Sean and Rosenberg, Kevin and Dawson, Scott TM and McKeon, Beverley J},
  journal={Physical Review Fluids},
  volume={3},
  number={5},
  pages={053902},
  year={2018},
  publisher={APS}
}

@article{jovanovic2021bypass,
  title={From bypass transition to flow control and data-driven turbulence modeling: an input--output viewpoint},
  author={Jovanovi{\'c}, Mihailo R},
  journal={Annual Review of Fluid Mechanics},
  volume={53},
  pages={311--345},
  year={2021},
  publisher={Annual Reviews}
}

@article{yeh2019resolvent,
  title={Resolvent-analysis-based design of airfoil separation control},
  author={Yeh, Chi-An and Taira, Kunihiko},
  journal={Journal of Fluid Mechanics},
  volume={867},
  pages={572--610},
  year={2019},
  publisher={Cambridge University Press}
}

@article{ribeiro2023triglobal,
  title={Triglobal resolvent analysis of swept-wing wakes},
  author={Ribeiro, JH Marques and Yeh, Chi-An and Taira, Kunihiko},
  journal={Journal of Fluid Mechanics},
  volume={954},
  pages={A42},
  year={2023},
  publisher={Cambridge University Press}
}

@article{franceschini2022identification,
  title={Identification and reconstruction of high-frequency fluctuations evolving on a low-frequency periodic limit cycle: application to turbulent cylinder flow},
  author={Franceschini, Lucas and Sipp, Denis and Marquet, Olivier and Moulin, Johann and Dandois, Julien},
  journal={Journal of Fluid Mechanics},
  volume={942},
  pages={A28},
  year={2022},
  publisher={Cambridge University Press}
}

@article{jovanovic2005componentwise,
  title={Componentwise energy amplification in channel flows},
  author={Jovanovi{\'c}, Mihailo R and Bamieh, Bassam},
  journal={Journal of Fluid Mechanics},
  volume={534},
  pages={145--183},
  year={2005},
  publisher={Cambridge University Press}
}

@article{mckeon2010critical,
  title={A critical-layer framework for turbulent pipe flow},
  author={McKeon, Beverley J and Sharma, Ati S},
  journal={Journal of Fluid Mechanics},
  volume={658},
  pages={336--382},
  year={2010},
  publisher={Cambridge University Press}
}

@article{liu2021unsteady,
  title={Unsteady control of supersonic turbulent cavity flow based on resolvent analysis},
  author={Liu, Qiong and Sun, Yiyang and Yeh, Chi-An and Ukeiley, Lawrence S and Cattafesta, Louis N and Taira, Kunihiko},
  journal={Journal of Fluid Mechanics},
  volume={925},
  pages={A5},
  year={2021},
  publisher={Cambridge University Press}
}

@article{rigas2021nonlinear,
  title={Nonlinear input/output analysis: application to boundary layer transition},
  author={Rigas, Georgios and Sipp, Denis and Colonius, Tim},
  journal={Journal of Fluid Mechanics},
  volume={911},
  pages={A15},
  year={2021},
  publisher={Cambridge University Press}
}

@inproceedings{wereley1990frequency,
  title={Frequency response of linear time periodic systems},
  author={Wereley, Norman M and Hall, Steven R},
  booktitle={29th IEEE conference on decision and control},
  pages={3650--3655},
  year={1990},
  organization={IEEE}
}

@Article{padovan2020analysis,
  title={Analysis of amplification mechanisms and cross-frequency interactions in nonlinear flows via the harmonic resolvent},
  author={Padovan, Alberto and Otto, Samuel E and Rowley, Clarence W},
  journal={Journal of Fluid Mechanics},
  volume={900},
  pages={A14},
  year={2020},
  publisher={Cambridge University Press}
}

@article{padovan2022analysis,
  title={Analysis of the dynamics of subharmonic flow structures via the harmonic resolvent: Application to vortex pairing in an axisymmetric jet},
  author={Padovan, Alberto and Rowley, Clarence W},
  journal={Physical Review Fluids},
  volume={7},
  number={7},
  pages={073903},
  year={2022},
  publisher={APS}
}

@book{garnier2009large,
  title={Large eddy simulation for compressible flows},
  author={Garnier, Eric and Adams, Nikolaus and Sagaut, Pierre},
  year={2009},
  publisher={Springer Science \& Business Media}
}

@article{chu1965energy,
  title={On the energy transfer to small disturbances in fluid flow (Part I)},
  author={Chu, Boa-Teh},
  journal={Acta Mechanica},
  volume={1},
  number={3},
  pages={215--234},
  year={1965},
  publisher={Springer}
}

@article{hanifi1996transient,
  title={Transient growth in compressible boundary layer flow},
  author={Hanifi, Ardeshir and Schmid, Peter J and Henningson, Dan S},
  journal={Physics of Fluids},
  volume={8},
  number={3},
  pages={826--837},
  year={1996},
  publisher={American Institute of Physics}
}

@article{bres2017unstructured,
  title={Unstructured large-eddy simulations of supersonic jets},
  author={Br{\`e}s, Guillaume A and Ham, Frank E and Nichols, Joseph W and Lele, Sanjiva K},
  journal={AIAA journal},
  volume={55},
  number={4},
  pages={1164--1184},
  year={2017},
  publisher={American Institute of Aeronautics and Astronautics}
}

@article{ribeiro2020randomized,
  title={Randomized resolvent analysis},
  author={Ribeiro, Jean H{\'e}lder Marques and Yeh, Chi-An and Taira, Kunihiko},
  journal={Physical Review Fluids},
  volume={5},
  number={3},
  pages={033902},
  year={2020},
  publisher={APS}
}

@techreport{rossiter1964wind,
  title={Wind-tunnel experiments on the flow over rectangular cavities at subsonic and transonic speeds},
  author={Rossiter, JE},
  institution={Aeronautical Research Council Reports and Memoranda},
   number ={3438},
  year={1964}
}

@article{sun2017biglobal,
  title={Biglobal instabilities of compressible open-cavity flows},
  author={Sun, Yiyang and Taira, Kunihiko and Cattafesta, Louis N and Ukeiley, Lawrence S},
  journal={Journal of Fluid Mechanics},
  volume={826},
  pages={270--301},
  year={2017},
  publisher={Cambridge University Press}
}

@article{toro2019hllc,
  title={The hllc riemann solver},
  author={Toro, Eleuterio F},
  journal={Shock waves},
  volume={29},
  number={8},
  pages={1065--1082},
  year={2019},
  publisher={Springer}
}

@article{leclercq2023mean,
  title={Mean resolvent operator of a statistically steady flow},
  author={Leclercq, Colin and Sipp, Denis},
  journal={Journal of Fluid Mechanics},
  volume={968},
  pages={A13},
  year={2023}
}

@article{freund1997proposed,
  title={Proposed inflow/outflow boundary condition for direct computation of aerodynamic sound},
  author={Freund, J. B.},
  journal={AIAA journal},
  volume={35},
  number={4},
  pages={740--742},
  year={1997}
}

@article{islam2024identification,
  title={Identification of cross-frequency interactions in compressible cavity flow using harmonic resolvent analysis},
  author={Islam, Md Rashidul and Sun, Yiyang},
  journal={Journal of Fluid Mechanics},
  volume={1000},
  pages={A13},
  year={2024},
  publisher={Cambridge University Press}
}

@book{jovanovic2004modeling,
  title={Modeling, analysis, and control of spatially distributed systems},
  author={Jovanovic, Mihailo R},
  year={2004},
  publisher={University of California, Santa Barbara}
}

@techreport{balay2019petsc,
  title={PETSc users manual},
  author={Balay, Satish and Abhyankar, Shrirang and Adams, Mark and Brown, Jed and Brune, Peter and Buschelman, Kris and Dalcin, Lisandro and Dener, Alp and Eijkhout, Victor and Gropp, William and others},
  year={2019},
  number={ANL-95/11- Revision 3.12},
  institution={Argonne National Laboratory}
}

@article{hernandez2005slepc,
  title={SLEPc: A scalable and flexible toolkit for the solution of eigenvalue problems},
  author={Hernandez, Vicente and Roman, Jose E and Vidal, Vicente},
  journal={ACM Transactions on Mathematical Software (TOMS)},
  volume={31},
  number={3},
  pages={351--362},
  year={2005},
  publisher={ACM New York, NY, USA}
}

@inproceedings{amestoy2000mumps,
  title={MUMPS: a general purpose distributed memory sparse solver},
  author={Amestoy, Patrick R and Duff, Iain S and L’Excellent, Jean-Yves and Koster, Jacko},
  booktitle={International Workshop on Applied Parallel Computing},
  pages={121--130},
  year={2000},
  organization={Springer}
}

@inproceedings{wereley1991linear,
  title={Linear time periodic systems: transfer function, poles, transmission zeroes and directional properties},
  author={Wereley, Norman M and Hall, Steven R},
  booktitle={1991 American Control Conference},
  pages={1179--1184},
  year={1991},
  organization={IEEE}
}

@article{sun2020resolvent,
  title={Resolvent analysis of compressible laminar and turbulent cavity flows},
  author={Sun, Yiyang and Liu, Qiong and Cattafesta III, Louis N and Ukeiley, Lawrence S and Taira, Kunihiko},
  journal={AIAA journal},
  volume={58},
  number={3},
  pages={1046--1055},
  year={2020},
  publisher={American Institute of Aeronautics and Astronautics}
}

@book{schmid2012stability,
  title={Stability and transition in shear flows},
  author={Schmid, Peter J and Henningson, Dan S},
  volume={142},
  year={2012},
  publisher={Springer Science \& Business Media}
}

@inproceedings{orr1907stability,
  title={The stability or instability of the steady motions of a perfect liquid and of a viscous liquid. Part II: A viscous liquid},
  author={Orr, William M'F},
  booktitle={Proceedings of the Royal Irish Academy. Section A: Mathematical and Physical Sciences},
  volume={27},
  pages={69--138},
  year={1907},
  organization={JSTOR}
}

@article{kachanov1984resonant,
  title={The resonant interaction of disturbances at laminar-turbulent transition in a boundary layer},
  author={Kachanov, Yu S and Levchenko, V Ya},
  journal={Journal of Fluid Mechanics},
  volume={138},
  pages={209--247},
  year={1984},
  publisher={Cambridge University Press}
}

@article{herbert1988secondary,
  title={Secondary instability of boundary layers},
  author={Herbert, Thorwald},
  journal={Annual review of fluid mechanics},
  volume={20},
  pages={487--526},
  year={1988}
}

@article{monokrousos2010global,
  title={Global three-dimensional optimal disturbances in the Blasius boundary-layer flow using time-steppers},
  author={Monokrousos, Antonios and {\AA}kervik, Espen and Brandt, Luca and Henningson, Dan S},
  journal={Journal of Fluid Mechanics},
  volume={650},
  pages={181--214},
  year={2010},
  publisher={Cambridge University Press}
}

@article{de2014three,
  title={Three-dimensional instabilities over a rectangular open cavity: from linear stability analysis to experimentation},
  author={de Vicente, Javier and Basley, Jeremy and Meseguer-Garrido, Fernando and Soria, Julio and Theofilis, Vassilios},
  journal={Journal of Fluid Mechanics},
  volume={748},
  pages={189--220},
  year={2014},
  publisher={Cambridge University Press}
}

@article{landahl1980note,
  title={A note on an algebraic instability of inviscid parallel shear flows},
  author={Landahl, MT},
  journal={Journal of Fluid Mechanics},
  volume={98},
  number={2},
  pages={243--251},
  year={1980},
  publisher={Cambridge University Press}
}

@article{bongarzone2025mean,
  title={Mean resolvent analysis of periodic flows},
  author={Bongarzone, Alessandro and Content, C{\'e}dric and Sipp, Denis and Leclercq, Colin},
  journal={arXiv preprint arXiv:2503.08401},
  year={2025}
}

@article{visbal2020passive,
  title={Passive control of dynamic stall using a flow-driven micro-cavity actuator},
  author={Visbal, M.R. and Garmann, D.J.},
  journal={Theor. Comput. Fluid Dyn},
  volume={37},
  pages={289--303},
  year={2023},
}

@article{rowley2006dynamics,
  title={Dynamics and control of high-Reynolds-number flow over open cavities},
  author={Rowley, Clarence W and Williams, David R},
  journal={Annu. Rev. Fluid Mech.},
  volume={38},
  number={1},
  pages={251--276},
  year={2006},
  publisher={Annual Reviews}
}

@article{rolandi2024invitation,
  title={An invitation to resolvent analysis},
  author={Rolandi, Laura Victoria and Ribeiro, Jean H{\'e}lder Marques and Yeh, Chi-An and Taira, Kunihiko},
  journal={Theoretical and Computational Fluid Dynamics},
  volume={38},
  number={5},
  pages={603--639},
  year={2024},
  publisher={Springer}
}

@article{Jung_Bhagwat_Towne_2025, 
title={Resolvent-based estimation and control of a laminar airfoil wake}, 
volume={1016},
DOI={10.1017/jfm.2025.10423}, journal={Journal of Fluid Mechanics}, author={Jung, Junoh and Bhagwat, Rutvij and Towne, Aaron}, year={2025}, pages={A41}}

@article{rowley2002self,
  title={On self-sustained oscillations in two-dimensional compressible flow over rectangular cavities},
  author={Rowley, Clarence W and Colonius, Tim and Basu, Amit J},
  journal={Journal of Fluid Mechanics},
  volume={455},
  pages={315--346},
  year={2002},
  publisher={Cambridge University Press}
}

@article{godavarthi2025phase,
  title={Phase-based analysis and control of supersonic turbulent cavity flows},
  author={Godavarthi, Vedasri and Kawamura, Yoji and Ukeiley, Lawrence S and Cattafesta III, Louis N and Taira, Kunihiko},
  journal={arXiv preprint arXiv:2502.05753},
  year={2025}
}

@inproceedings{victorino2023gap,
  title={Gap induced boundary layer transition},
  author={Victorino, Victor B and Aguirre, Felipe O and Medeiros, Marcello A},
  booktitle={AIAA AVIATION 2023 Forum},
  pages={3997},
  year={2023}
}

@article{heller1971flow,
  title={Flow-induced pressure oscillations in shallow cavities},
  author={Heller, Hanno H and Holmes, DG and Covert, Eugene E},
  journal={Journal of sound and Vibration},
  volume={18},
  number={4},
  pages={545--553},
  year={1971},
  publisher={Elsevier}
}
\end{document}